\documentstyle[11pt,epsf]{article}
\begin{document}
\def\mathrm#1{{\rm #1}}

\hyphenation{meta-stable}


\title{\bf
Application of a constrained-transfer-matrix method to metastability in
the $d$$=$$2$ Ising ferromagnet\thanks{Submitted to {\it Physica A}.}
}

\author{C.~C.~A.~G{\"u}nther,$^{\rm a,b,{\scriptsize\dag},}$\thanks{Partially
  supported by Florida State University through
  the Supercomputer Computations Research Institute (U.S.~Department
  of Energy Contract No.~DE-FC05-85ER25000) and through Cray Y-MP
  supercomputer time.}\thanks{Partially supported by the
  Florida State University
  Center for Materials Research and Technology and by U.S.~National
  Science Foundation Grants No.~DMR-9013107 and DMR-9315969.}
{}~P.~A.~Rikvold,$^{\rm a,b,{\scriptsize\dag},{\scriptsize\ddag}}$
M.~A.~Novotny$^{\rm a,{\scriptsize\dag}}$\\[1ex]
$^{\rm a}$\  Supercomputer Computations Research Institute,\\
Florida State University, Tallahassee, Florida 32306-4052, USA\\[1ex]
$^{\rm b}$\ Department of Physics and Center for Materials
 Research and Technology,\\
 Florida State University, Tallahassee, Florida 32306-3016, USA
}
\date{}

\maketitle


\begin{abstract}
\typeout{abstract}
Applying a numerical transfer-matrix formalism, we obtain
complex-valued constrained free energies for the two-dimensional
square-lattice nearest-neighbor Ising ferromagnet below its critical
temperature and in an external magnetic field. In particular, we
study the imaginary part of the constrained free-energy branch that
corresponds to the metastable phase. Although droplets are not
introduced explicitly, the metastable free energy is obtained in
excellent agreement with field-theoretical droplet-model predictions.
The finite-size scaling properties are different in the weak-field
and intermediate-field regimes, and we identify the corresponding
different critical-droplet shapes. For intermediate fields, we show
that the surface free energy of the critical droplet is given by a
Wulff construction with the equilibrium surface tension. We also find
a prefactor exponent in complete agreement with the field-theoretical
droplet model. Our results extend the region of validity for known
results of this field-theoretical droplet model, and they indicate
that this transfer-matrix approach provides a nonperturbative
numerical continuation of the equilibrium free energy into the
metastable phase.
\end{abstract}




\section{Introduction}
\typeout{1. Introduction}
\label{sec1}
Metastability is commonly observed in a wide variety of systems,
ranging from supercooled fluids and vapors
\cite{mcd62,mcd63,abr74,oxt92} to the electroweak
\cite{gle92,kri92,buc92} and QCD confinement/deconfinement
\cite{kaj92,hac92} phase transitions, and its description in terms
of statistical mechanics has received considerable attention. (See
e.g.\ Refs.~\cite{oxt92,gun83,gun83i,bin87,rik94i} for reviews.)
Nevertheless, a fully satisfactory microscopic description remains
elusive \cite{schu90}. One characteristic feature of metastable
phases is that although they do not fully minimize the free energy,
they nevertheless display long-term stability against small
perturbations, with lifetimes that can be many orders of magnitude
longer than other characteristic timescales of the system
\cite{pen71,pen79,sew80}. Based on this observation, numerous efforts
to formulate a statistical theory of metastability have treated
metastable phases as ``quasi-equilibrium'' phases derived from a
constrained partition function that excludes or severely reduces the
probabilities of those microstates that are more probable in
equilibrium. In the context of a field-theoretical droplet model
with Fokker-Planck dynamics, these ideas led to the result
\cite{lan67,lan68,lan69} that the nucleation rate of a metastable
phase is proportional to the imaginary part of the analytic
continuation of the equilibrium free energy into the metastable
phase. However, the original derivation of this result is only valid
for large systems in the limit of ultraweak magnetic fields
\cite{lan68,lan69}, and despite extensive subsequent studies
\cite{schu90,new77,mcc78,pri82a,pri82b,new80,roe84,gav87,gav89}, its
domain of validity still remains unclear.

In one of these subsequent studies \cite{new77}, it was suggested
that the analytic continuation of the Ising model free energy could
be found from certain analytic properties of the eigenvalues of the
transfer matrix. Further transfer-matrix work performed on the
two-dimensional nearest-neighbor Ising model
\cite{mcc78,pri82a,pri82b} supported this result. Furthermore,
these studies found indications for the existence of an essential
singularity at zero magnetic field, as expected from field-theoretical
droplet-model calculations
\cite{lan67,lan68,lan69,and64,fis67,gue80,zia81}. Since then, this
result has been confirmed by series expansions
\cite{low80,bak80,bax79,ent80,har84}, exact diagonalization studies
\cite{ham83}, Monte Carlo simulations \cite{jac83,per84}, rigorous
studies using low-temperature Peierls contours \cite{isa84}, and
calculations of two-point correlation functions in a bubble model
\cite{abr92,abr92i,abr93}.

Motivated by the transfer-matrix results mentioned above, one of us
\cite{rik89,rik91a} has introduced a ``constrained-transfer-matrix''
(CTM) method. It generalizes conventional transfer-matrix (TM)
techniques, providing a means to compute complex-valued ``constrained
free energies'' from the eigenvalues and eigenvectors of the transfer
matrix. Here we report in detail on the application of the CTM method
to a short-range-force model, the two-dimensional ferromagnetic
nearest-neighbor Ising model on a square lattice. We find that the
imaginary part of the constrained free energy that corresponds to
the metastable phase, as obtained by the CTM method, is in excellent
agreement with predictions of the field-theoretical droplet model. A
brief account of some aspects of this work has been published
elsewhere \cite{gue93}.

One alternative approach to the study of metastability is to perform
Monte Carlo simulations and measure the lifetime of the metastable
phase directly. For the two-dimensional nearest-neighbor Ising
ferromagnet this was done in
Refs.~\cite{mcc78,sto72,bin73a,bin74,sto77,bin80,pau88} and more
recently in Refs.~\cite{dui90,ray90a,tom92,rik94,leex,nov94,novx}.
We compare this lifetime with the imaginary part of the metastable
free energy as obtained by the CTM method and find very similar
dependences on the magnetic field, the temperature, and the finite
size of the system. This strongly supports the validity of the
aforementioned proportionality relation between the nucleation rate
and the imaginary part of the metastable free energy.

A comparison of the CTM formalism applied to a short-range-force
model, as presented here, and to two- and three-state
models with weak long-range forces, presented elsewhere
\cite{rik89,rik91a,rik92a,gor94a,fii94}, clearly reveals the
differences and similarities between these models \cite{gor94b}. The
analytic continuation of the equilibrium free energy in
long-range-force models has a branch-point singularity at a
well-defined non-zero spinodal field
\cite{schu90,pen71,pen79,sew80,roe84,gav87,gav89,rik89,rik91a,rik92a,gor94a,fii94},
whereas it has an essential singularity on the coexistence line at
zero magnetic field in short-range-force models
\cite{and64,fis67,low80,bak80,bax79,ent80,har84,ham83,jac83,per84,isa84,abr92,abr92i,abr93}.
Moreover, whereas long-range-force models exhibit infinitely
long-lived metastable phases in the limit of infinite interaction
range, short-range-force models display finite, albeit very long
lifetimes, even in the thermodynamic limit
\cite{mcc78,sto72,bin73a,bin74,sto77,bin80,pau88,dui90,ray90a,tom92,rik94,leex,nov94,novx,sta82,hee84,ray90c,sta92,bea94}.
However, the results for the imaginary part of the metastable free
energy in all of these cases can be explained by assuming that the
rate-determining step in the decay process is the creation of a
critical excitation whose free-energy cost can be considered as an
activation energy.

Our results lead to the conclusion that the CTM formalism provides
a method to numerically continue the equilibrium free energy into
the metastable region. In contrast to the analytic continuation
in field-theoretical droplet models, the CTM formalism does not
rely on the explicit introduction of droplets.

The remainder of this article is organized as follows. In
Sec.~\ref{sec2} we explain the CTM method and describe the numerical
methods used to calculate the eigenvalues and eigenvectors of the TM.
Section \ref{sec3} gives a brief overview of the droplet theory with
emphasis on those aspects that are relevant to the interpretation of
our results. The numerical results and their interpretation are
presented in Sec.~\ref{sec4}, followed by a summary and conclusions
in Sec.~\ref{sec5}.


\section{The Constrained-Transfer-Matrix Method}
\typeout{2. The Constrained-Transfer-Matrix Method}
\label{sec2}


\subsection{Theory}
\typeout{2.1. Theory}
\label{subsec2.1}
We consider the two-dimensional nearest-neighbor Ising ferromagnet
on an $N$$\times$$L$ lattice at a temperature below its critical
temperature $T_{{\mathrm{c}}}$. The Hamiltonian is given by
\begin{equation}
{\cal H}=-J \sum_{\langle i,j \rangle} s_i s_j
                    - H \sum_i s_i \; ,         \label{eq2.1}
\end{equation}
where $s_i$$=$$\pm 1$ denotes the spin at site $i$, $J$$>$$0$
is the nearest-neighbor coupling, the sums run over all
nearest-neighbor pairs and all sites, respectively, and $H$ is
the externally applied magnetic field. Throughout this work we use
periodic boundary conditions.

The conventional TM method is a technique to calculate standard
thermodynamic state functions in equilibrium systems, such as the
magnetization, the internal energy, the free energy, and correlation
functions \cite{dom60}. In order to define the TM, one first divides
the $N$$\times$$L$ lattice into $L$ one-dimensional subsystems of
length $N$. A subsystem configuration $|x_l \rangle$ can be written
as a direct product of single-spin configurations in that subsystem:
$|x_l \rangle$$\equiv$$|s_{1,l} \rangle \cdots |s_{N,l} \rangle$.
One then decomposes the total Hamiltonian (e.g.\ Eq.~(\ref{eq2.1}))
into a sum of subsystem Hamiltonians $\bar{\cal H}$:
\begin{equation}
{\cal H} = \sum_{l=1}^L \bar{\cal H}(x_l,x_{l+1}) \; .
                                                     \label{eq2.2}
\end{equation}
The elements of the $2^N$$\times$$2^N$ transfer matrix (TM)
${\bf T}_0$ are then given by
\begin{equation}
\langle x_l |{\bf T}_0| x_{l+1} \rangle =
                         \exp[-\beta \bar{\cal H}(x_l,x_{l+1})] \; .
                                     \label{eq2.3}
\end{equation}
Here $\beta$$=$$1/T$ is the inverse temperature with Boltzmann's
constant set equal to unity. Using Eqs.~(\ref{eq2.1}), (\ref{eq2.3}),
and periodic boundary conditions in the $L$ direction, the equilibrium
partition function can be written as
\begin{equation}
Z=\mathrm{Tr}\left({\bf T}_0^L\right)=\sum_\alpha \lambda^L_\alpha \; ,
                                          \label{eq2.4}
\end{equation}
where $\lambda_\alpha$, $0$$\,\leq\,$$\alpha$$\,\leq\,$$2^N-1$,
denote the eigenvalues of the TM in order of decreasing magnitude.
Since ${\bf T}_0$ is finite with all positive elements, the
Perron--Frobenius theorem states that the eigenvalue of largest norm,
$\lambda_0$, is positive and nondegenerate \cite{dom60}. Therefore, in
the limit $L$$\rightarrow$$\infty$, the equilibrium partition
function is given by
\begin{equation}
Z=\lambda_0^L \; .   \label{eq2.5}
\end{equation}
For the equilibrium free-energy density, $f_0$, we then obtain
\begin{equation}
f_0 = -\frac{1}{\beta N}\ln \lambda_0 \; .   \label{eq2.6}
\end{equation}
The equilibrium joint and marginal probability densities are given by
\cite{dom60}
\begin{equation}
P_0(x_l,x_{l+k}) = \langle 0|x_l \rangle \langle
                   x_l|(\lambda_0^{-1}{\bf T}_0)^{|k|}|x_{l+k} \rangle
                   \langle x_{l+k}|0 \rangle  \label{eq2.7}
\end{equation}
and
\begin{equation}
P_0(x_l)=\langle 0|x_l \rangle \langle x_l|0 \rangle \; . \label{eq2.8}
\end{equation}
In terms of the joint and marginal probability densities, the
equilibrium entropy is given by
\begin{equation}
S_0 = -\frac{1}{N} \sum_{x_l,x_{l+1}} P_0(x_l,x_{l+1}) \ln \left[
       \frac{P_0(x_l,x_{l+1})}{P_0(x_l)} \right] \; .   \label{eq2.9}
\end{equation}

The CTM formalism \cite{rik89,rik91a} extends this scheme to
nonequilibrium states by defining constrained probability densities
using the orthonormal eigenvectors $|\alpha \rangle$ that correspond
to the subdominant eigenvalues of ${\bf T}_0$:
\begin{eqnarray}
P_{\alpha}(x_l,x_{l+k}) & = & \langle \alpha|x_l \rangle \langle x_l
                     |(\lambda_{\alpha}^{-1} {\bf T}_{\alpha})^{|k|}
                     |x_{l+k} \rangle \langle x_{l+k}|\alpha \rangle
                                            \label{eq2.10}  \\
P_{\alpha}(x_l) & = & \langle \alpha| x_l \rangle \langle x_l
                               |\alpha \rangle \; .
                                            \label{eq2.11}
\end{eqnarray}
The marginal probability densities, Eq.~(\ref{eq2.11}), are
normalized and can be interpreted as probability densities of
subsystem configurations in a constrained state
\cite{new77,mcc78,pri82a,pri82b,new80}. It is apparent from
Eqs.~(\ref{eq2.7}), (\ref{eq2.8}), (\ref{eq2.10}), and (\ref{eq2.11})
that the constrained probability densities reduce to their equilibrium
counterparts for $\alpha$$=$$0$.

For simplicity, we restrict ourselves in the following discussion to
symmetric ${\bf T}_0$ and assume that all its eigenvalues are
nondegenerate. (In a large number of cases of physical interest,
these conditions are or can be made to be satisfied.) The matrices
${\bf T}_\alpha$ are chosen to commute with ${\bf T}_0$ and can hence
be expanded in the eigenvectors, $| \beta \rangle$, of ${\bf T}_0$:
\begin{equation}
{\bf T}_\alpha = \sum_\beta |\beta \rangle \mu_\beta(\alpha)
                  \langle \beta|.            \label{eq2.12}
\end{equation}

The ``reweighted'' eigenvalues $\mu_\beta(\alpha)$ are chosen so
that ${\bf T}_\alpha$, for $\alpha$$>$$0$, has $\lambda_\alpha$
as its dominant eigenvalue, rather than $\lambda_0$. Furthermore,
in order for the entire multi-layer system to be in a uniform state
consistent with $P_\alpha(x_l)$, the joint probability densities
$P_\alpha(x_l,x_{l+k})$ have to fulfill the following requirements
\cite{rik89}:
\begin{enumerate}
\item Stochastic independence at large separation:
\begin{equation}
\lim_{|k| \rightarrow \infty}P_\alpha(x_l,x_{l+k})
             = P_\alpha(x_l) P_\alpha(x_{l+k}) \; ,    \label{eq2.13}
\end{equation}
which ensures that fluctuations corresponding to eigenstates
orthogonal to $|\alpha \rangle$ decay on a finite length scale. This
condition is satisfied if and only if
$|\mu_\beta(\alpha)|$$<$$|\lambda_\alpha|$ for all
$\beta$$\neq$$\alpha$.
\item Standard relations between the joint and marginal probability
densities:
\begin{equation}
\sum_{x_{l+k}} P_\alpha(x_l,x_{l+k}) = P_\alpha(x_l) \; ,
                                                    \label{eq2.14}
\end{equation}
which is satisfied if and only if
$\mu_\beta(\alpha)$$=$$\lambda_\alpha$ for $\beta$$=$$\alpha$.
\item Non-ambiguity of Eqs.~(\ref{eq2.10}) and (\ref{eq2.11}) for
$k$$=$$0$:
\begin{equation}
P_\alpha(x_l,x_l^\prime) = P_\alpha(x_l) \delta_{x_l,x_l^\prime} \; .
                                                    \label{eq2.15}
\end{equation}
This condition requires the matrices ${\bf T}_\alpha$ to be of the
same rank as ${\bf T}_0$. It is satisfied if and only if, for all
$\alpha$, $\mu_\beta(\alpha)$$\neq$$0$ for all $\beta$ for which
$\lambda_\beta$$\neq$$0$.
\end{enumerate}
These requirements, however, do not uniquely give the detailed
form of ${\bf T}_{\alpha}$. Since all the
$\lambda_{\alpha}$$\neq$$0$ for the system studied here, a
reweighting scheme that satisfies Eqs.~(\ref{eq2.13}),
(\ref{eq2.14}), and (\ref{eq2.15}), and which was also chosen in
Refs.~\cite{rik89,rik91a,gue93,rik92a,gor94a,fii94}, is given by
\begin{equation}
\mu_\beta(\alpha) = \left\{ \begin{array}{ll}
                      \lambda_\alpha^2/\lambda_\beta &
                      \mbox{if $|\lambda_\beta| > |\lambda_\alpha|$} \\
                      \lambda_\beta &
                      \mbox{if $|\lambda_\beta| \leq |\lambda_\alpha|$}
                      \end{array}
                    \right. \! .    \label{eq2.16}
\end{equation}
Note that the ${\bf T}_\alpha$ for $\alpha$$>$$0$ are
in general {\em not\/} positive matrices. Since only the eigenvector
$|0 \rangle$, corresponding to the largest eigenvalue, can be chosen
to have all non-negative elements \cite{dom60}, the other
eigenvectors, which are orthogonal to $|0 \rangle$, necessarily contain
some negative and some positive elements. Therefore, at least some of
the matrices ${\bf T}_\alpha$ for $\alpha$$>$$0$ may not be
positive.

In order to define ``constrained'' state functions, it is convenient
to decompose the subsystem Hamiltonian, $\bar{\cal H}(x_l,x_{l+1})$,
into two parts: $\bar{\cal H}_{\mathrm{I}}$, containing only
interaction terms, and $-H \bar{\cal M}$, containing only terms
proportional to the magnetic field. A constrained internal energy
per site, $U_\alpha$, and a constrained field energy, $-H M_\alpha$,
can then be defined by replacing the equilibrium probability
densities in the expression for the equilibrium expectation values of
these quantities with the constrained probability densities. One
obtains
\begin{equation}
U_\alpha = \frac{1}{N} \sum_{x_l,x_{l+1}} P_\alpha(x_l,x_{l+1})
                              \bar{\cal H}_{\mathrm{I}}(x_l,x_{l+1})
                                                       \label{eq2.17}
\end{equation}
and
\begin{equation}
-H M_\alpha = -\frac{H}{N} \sum_{x_l,x_{l+1}} P_\alpha(x_l,x_{l+1})
                                     \bar{\cal M}(x_l,x_{l+1}) \; .
                                                     \label{eq2.18}
\end{equation}
Constrained entropy densities $S_\alpha$ are defined in analogy with
the source entropy of a stationary, ergodic Markov information source
(see e.g.\ \cite{bla87}):
\begin{eqnarray}
S_\alpha & = & -\frac{1}{N} \sum_{x_l,x_{l+1}} P_\alpha (x_l,x_{l+1}) \,
     \mathrm{Ln} \left[ \frac{P_\alpha (x_l,x_{l+1})}{P_\alpha (x_l)}
                   \right]   \label{eq2.19} \\
         & = & -\frac{1}{N} \sum_{x_l,x_{l+1}} P_\alpha (x_l,x_{l+1}) \,
  \mathrm{Ln} \langle x_l |\lambda_\alpha^{-1} {\bf T}_\alpha| x_{l+1}
                   \rangle \; .               \nonumber
\end{eqnarray}
Since the elements of ${\bf T}_\alpha$ are in general not
non-negative,  this gives rise to an imaginary part in $S_\alpha$
through the principal branch of the complex logarithm
$\mathrm{Ln}z$$=$$\ln|z|$$+$$\mathrm{i}\phi$ with
$z$$=$$|z|\mathrm{e}^{\mathrm{i} \phi}$ and
$-\pi$$\leq$$\phi$$<$$\pi$. Notice also that for
$\alpha$$=$$0$, $U_\alpha$, $M_\alpha$, and $S_\alpha$ reduce
to the corresponding equilibrium quantities.

In analogy with equilibrium thermodynamics, constrained free-energy
densities, $f_\alpha$, are then defined by
\begin{equation}
f_\alpha = U_\alpha - H M_\alpha - \beta^{-1} S_\alpha \; ,
                                  \label{eq2.20}
\end{equation}
which may also be written as
\begin{equation}
f_\alpha = -\frac{1}{\beta N} \ln |\lambda_\alpha|
        +\frac{1}{\beta N} \sum_{x_l,x_{l+1}} P_\alpha (x_l,x_{l+1}) \,
        \mathrm{Ln} \left[
                \frac{\langle x_l |{\bf T}_\alpha| x_{l+1} \rangle}
                {\langle x_l |{\bf T}_0| x_{l+1} \rangle} \right] \; .
                                          \label{eq2.21}
\end{equation}
The first term is analogous to the equilibrium case, whereas the
second term vanishes in equilibrium. Moreover, the second term is
in general complex-valued, since the ${\bf T}_\alpha$ are not
necessarily positive matrices. Formally, this second term can be
considered as a complex generalization of the Kullback discrimination
function (see e.g.\ \cite{kap92}) for $P_\alpha (x_l,x_{l+1})$ with
respect to the divergent `probability density' obtained by
substituting ${\bf T}_0$ for ${\bf T}_\alpha$ in Eq.~(\ref{eq2.10}).


\subsection{Numerical methods}
\typeout{2.2. Numerical method}
\label{subsec2.2}
In order to calculate any of the constrained free-energy densities
in Eq.~(\ref{eq2.20}), one needs to know {\it all} of the $2^N$
eigenvalues and eigenvectors of the equilibrium transfer matrix
${\bf T}_0$. The attainable system sizes can be increased
considerably by making use of the symmetries of ${\bf T}_0$ (see
e.g.\ \cite{nig90}). Let $X$ be the $2^N$-dimensional vector space
with the basis $\{ |x_j \rangle \}$ ($j=1,2, \ldots , 2^N$) of
subsystem configurations. The symmetries of ${\bf T}_0$ are
represented by unitary operators ${\bf U}:X \rightarrow X$ that
commute with ${\bf T}_0$.  Applying ${\bf U}$ to a particular
configuration $|x_j \rangle$ amounts to a permutation of the spins
in $|x_j \rangle$. The application of ${\bf U}$ to the configurations
in $X$ partitions $X$ into subsets, each of which is invariant under
${\bf U}$. Since the number $N$ of spins within a subsystem is
finite, ${\bf U}$ has a finite period $M$$\leq$$N$ with
${\bf U}^M$$=$${\bf I}$, where ${\bf I}$ is the identity in $X$.
This means that the eigenvalues of ${\bf U}$ are given by the $M$
roots of unity. Each eigenvalue of ${\bf U}$ corresponds to one of
the subsets of $X$.

One can now block-diagonalize ${\bf T}_0$ by writing it in the
eigenvector basis of ${\bf U}$, with each block corresponding to a
different eigenvalue of ${\bf U}$. The blocks corresponding to the
eigenvalues $1$ and $-1$ are called symmetric and antisymmetric,
respectively. They are real and symmetric, whereas all the other
blocks are Hermitian. Instead of diagonalizing the full
$2^N \times 2^N$ transfer matrix all at once, one now diagonalizes
each block separately, with the submatrices corresponding to the
Hermitian blocks first converted into real symmetric matrices
\cite{numrec}.

In our case, ${\bf T}_0$ is symmetric under cyclic permutations
${\bf P}$ and reflections ${\bf R}$ of the subsystem configurations.
Since ${\bf P}$ has a period $N$ that in general is larger than the
period of ${\bf R}$, we block-diagonalize ${\bf T}_0$ with respect
to ${\bf P}$. In order to identify those states that contribute to
the metastable phase, we reduce the symmetric block of ${\bf T}_0$
further by block-diagonalizing it again with respect to ${\bf R}$.
This exploits the fact that the metastable phase has the same
symmetry as the stable phase, which is symmetric under both ${\bf P}$
and ${\bf R}$ \cite{mcc78,pri82a,pri82b}.

The diagonalization is performed in two steps. First, we reduce the
matrices to tridiagonal form by the Householder method, and then we
use a QL algorithm with implicit shifts to find the eigenvectors and
eigenvalues of the resulting tridiagonal matrices \cite{numrec}.

Finally, in order to calculate the constrained free-energy densities
we transform the eigenvectors of the transfer matrix, which are now
written in terms of the basis in which ${\bf T}_0$ is block-diagonal,
back into the basis $\{ |x_j \rangle \}$ ($j=1, \ldots 2^N$) of
subsystem configurations. The unitary transformation to accomplish
this is constructed from the eigenvectors of ${\bf P}$ and ${\bf R}$
\cite{nov86p}.

It turns out that the imaginary parts of the metastable constrained
free-energy densities are extremely small (as will be seen later in
Fig.~3(b)), although the magnitude of each individual term
in the sum of Eq.~(\ref{eq2.21}) may be on the order of unity. In
order to attain sufficient precision, the diagonalization was
therefore performed in 128-bit precision on a Cray Y-MP/432 vector
supercomputer. The total computer time spent in this study was on the
order of 1000 CPU hours.

For temperatures $T/J$$\leq$$0.4$, the limit on the attainable
system sizes is given by numerical underflow for small $|H|$,
preventing us from studying system sizes larger than $N$$=$$9$
for $T/J$$=$$0.4$. For all higher temperatures studied, the
maximum system size attainable is $N$$=$$10$, limited by the
available computer memory.


\section{Droplet Theory}
\typeout{3. Droplet Theory}
\label{sec3}
Suppose we prepare the system with $H$$<$$0$, so that its average
magnetization is close to $-1$, and then quench the field
through $H$$=$$0$ to a value $H$$>$$0$, such that the average
magnetization is approximately unchanged. This state is then no
longer a global minimum of the free energy. Nevertheless, it might
persist for a period of time many orders of magnitude longer than
other characteristic timescales of the system
\cite{pen71,pen79,sew80}. This can be used as an operational
definition of metastability.

Starting with the Becker--D\"{o}ring droplet model
\cite{gun83,gun83i}, one assumes that the decay of metastable phases
proceeds through the spontaneous formation of droplets of up spins in
the background of down spins. Assuming further that the distance
between droplets is much larger than their characteristic size (which
is reasonable for sufficiently low temperatures and weak fields
\cite{rik94,sek86,sek91}), one can treat them as a gas of
noninteracting droplets. (In the following, we always assume that $H$
is sufficiently weak, so that interactions between droplets are
negligible.) The probability of creating a droplet is then
proportional to the Boltzmann factor
$\mathrm{e}^{-\beta F_{\mathrm{D}}}$, where $F_{\mathrm{D}}$ is the
free energy of the droplet. The final assumption of the
Becker--D\"{o}ring droplet model is that $F_{\mathrm{D}}$ is given by
the sum of the surface free energy and the bulk free energy of the
droplet. If the surface free energy does not change with $H$ and is
an increasing and convex function of the volume $V$ of the droplet,
and if the bulk free energy is a decreasing function of $H$ and $V$
then a critical droplet with volume $V_{\mathrm{c}}$ exists so that
droplets with $V$$>$$V_{\mathrm{c}}$ tend to grow, leading to the
decay of the metastable phase. It also follows that $V_{\mathrm{c}}$
increases with decreasing $H$. The quantity $V_{\mathrm{c}}$ can be
obtained by maximizing $F_{\mathrm{D}}$.


\subsection{Infinite systems}
\typeout{3.1. Infinite systems}
\label{subsec3.1}
We assume for the moment that the system is infinitely large. We then
express the surface free energy, $\Sigma$, in the form
$\Sigma$$=$$V^{\frac{d-1}{d}} \widehat{\Sigma}$, where $d$ is
the spatial dimension, and we denote the difference between the
bulk free-energy densities of the metastable and the stable phase
by $\Delta f$. Then $F_{\mathrm{D}}$ can be written as
\begin{equation}
F_{\mathrm{D}} = V^{\frac{d-1}{d}} \widehat{\Sigma}-V \Delta f \; .
                                    \label{eq3.1}
\end{equation}
Maximizing $F_{\mathrm{D}}$ yields
\begin{equation}
V_{\mathrm{c}} = \left( \frac{(d-1) \widehat{\Sigma}}{d \Delta f}
            \right)^d \; .
                                                  \label{eq3.2}
\end{equation}

Let $\Delta m$$\equiv$$m_{{\mathrm{ms}}} - m_{{\mathrm{st}}}$,
where $m_{{\mathrm{ms}}}$ and $m_{{\mathrm{st}}}$ denote the
magnetizations of the metastable and stable phases, respectively.
The approximation $\Delta f$$\approx$$|H| \Delta m$ incorporates
the effects of droplet nesting, which becomes increasingly important
as the temperature increases from zero to $T_{\mathrm{c}}$
\cite{har84}. Inserting $|H| \Delta m$ and Eq.~(\ref{eq3.2}) into
Eq.~(\ref{eq3.1}), the free-energy cost of the critical droplet,
$F_{\mathrm{c}}(T,H)$, is obtained as
\begin{equation}
F_{\mathrm{c}}(T,H) = \left( \frac{\widehat{\Sigma}}{d} \right)^{d}
                  \left( \frac{d-1}{|H| \Delta m} \right)^{d-1} \; .
                                       \label{eq3.3}
\end{equation}

Within the context of the droplet model, the nucleation rate of
critical droplets per unit volume is proportional to the probability
of creating a critical droplet and can hence be written in the form
\cite{gun83,gun83i,lan68,lan69}
\begin{equation}
I(T,H) = I_0(T,H) \mathrm{e}^{-\beta F_{\mathrm{c}}(T,H)}
       = I_0(T,H) \exp \left(- \frac{\beta \Xi}{|H|^{d-1}} \right)
                                      \label{eq3.4}
\end{equation}
with
\begin{equation}
\Xi = \left( \frac{\widehat{\Sigma}}{d} \right)^d
      \left( \frac{d-1}{\Delta m} \right)^{d-1} \; .
                                      \label{eq3.5}
\end{equation}

The proportionality factor $I_0(T,H)$ in Eq.~(\ref{eq3.4}) can be
derived by going beyond the Becker--D\"{o}ring droplet model to a
field-theoretical droplet model with Fokker-Planck dynamics. Within
this framework it has been shown \cite{lan67,lan68,lan69} that for
infinitely large systems in ultraweak fields, the nucleation rate
per unit volume is proportional to the imaginary part of a
complex-valued ``constrained'' free-energy density,
Im$f_{\mathrm{ms}}$, obtained by analytic continuation of the
equilibrium free energy into the metastable phase \cite{lan68,lan69}:
\begin{equation}
I(T,H) = \frac{\beta \kappa}{\pi} |\mathrm{Im} f_{\mathrm{ms}}|
                                                   \; . \label{eq3.6}
\end{equation}
Here $\kappa$ is a kinetic prefactor which depends on the details
of the dynamics, and the imaginary part of the constrained
free-energy density is given by \cite{lan67,gue80}
\begin{equation}
|\mathrm{Im} f_{\mathrm{ms}}| = B(T) |H|^b
                        \exp \left(-\beta F_{\mathrm{c}}(T,H)
                             \left[ 1+ O(H^2) \right] \right) \; ,
                                      \label{eq3.7}
\end{equation}
Inserting $F_{\mathrm{c}}(T,H)$ from Eq.~(\ref{eq3.3}) into
Eq.~(\ref{eq3.7}), one obtains
\begin{equation}
|\mathrm{Im} f_{\mathrm{ms}}| = B(T) |H|^b \exp \left(-\frac{\beta \Xi}
            {|H|^{d-1}}\left[ 1+ O(H^2) \right] \right) \; ,
                                      \label{eq3.7a}
\end{equation}
where $\Xi$ is given by Eq.~(\ref{eq3.5}). The function $B(T)$ is
expected to be nonuniversal, whereas the exponent $b$ is universal
and related to surface excitations (Goldstone modes) of the critical
droplet \cite{lan67,gue80}:
\begin{equation}
b = \left\{ \begin{array}{ll}
             (3-d)d/2 & \mbox{\ \ for \ $1$$<$$d$$<$$5$, \
                                          $d$$\neq$$3$} \\
             -7/3     & \mbox{\ \ for \ $d$$=$$3$}
           \end{array}
    \right. \; .                       \label{eq3.8}
\end{equation}
Omitting the surface excitations in the derivation of
Eq.~(\ref{eq3.7a}) leads to \cite{gorpc}
\begin{equation}
b = -1 - \frac{(d-1)d}{2} \; .            \label{eq3.9}
\end{equation}
There is strong numerical evidence that $b$$=$$1$ for $d$$=$$2$
\cite{low80,har84,rik94,wal80}, in agreement with Eq.~(\ref{eq3.8})
and disagreement with Eq.~(\ref{eq3.9}).

Under the assumption that the critical droplet is sufficiently
compact, one can define its ``radius'' $R_{\mathrm{c}}$ as half of
its spanning length along one of the primitive lattice vectors. One
can then write
\begin{equation}
V_{\mathrm{c}}= \Omega_d R_{\mathrm{c}}^d \; ,      \label{eq3.10}
\end{equation}
where $\Omega_d$ incorporates the detailed shape of the droplet.
Using Eq.~(\ref{eq3.2}) this leads to
\begin{equation}
R_{\mathrm{c}} = \frac{(d-1)\widehat{\Sigma}}{d\Omega_d^{1/d}\Delta f}
                                     \; . \label{eq3.11}
\end{equation}

Making the additional assumption that the critical droplet shape can
be obtained from a Wulff construction with an anisotropic surface
tension, one can relate the shape factor $\Omega_d$ to the surface
tension. Let $\sigma(\hat{n})$ be the anisotropic surface tension
of an interface with normal $\hat{n}$. In the Wulff construction one
first draws a polar plot of the surface tension. According to Wulff's
theorem, the shape of the droplet is then given by the inner envelope
of the geometric construction obtained by drawing a line tangential
to the polar plot of $\sigma(\hat{n})$ for each direction $\hat{n}$,
i.e.\ \cite{rot81,zia82,avr82,rot84,wor88}
\begin{equation}
\lambda R(\hat{r}) = {\mathrm{min}}_{\hat{n}}
                  \frac{\sigma(\hat{n})}{\hat{n} \cdot \hat{r}} \; .
                             \label{eq3.12}
\end{equation}
Here $R(\hat{r})$ is the length of the radius vector in the
direction $\hat{r}$, and $\lambda$ is a scale factor to be determined
from the actual size of the droplet.

If $W$ denotes the volume bounded by $R(\hat{r})$, as given by
Eq.~(\ref{eq3.12}) with $\lambda$$=$$1$, then the total surface
free energy of the droplet is given by $\Sigma$$=$$d W^{1/d}
V^{(d-1)/d}$ \cite{zia82}. With $\Sigma$$=$$\widehat{\Sigma}
V^{(d-1)/d}$, this leads to
\begin{equation}
W = \left( \frac{\widehat{\Sigma}}{d} \right)^d \; .   \label{eq3.13}
\end{equation}
Furthermore, in the Wulff construction with $\lambda$$=$$1$, the
length of the radius vector of $W$ in any direction is the surface
tension in that direction. Therefore $W$ and $V_{\mathrm{c}}$ are
related by $V_{\mathrm{c}}$$=$$(R_{\mathrm{c}}/\sigma_0)^d W$,
where $\sigma_0$ denotes the surface tension in one of the symmetry
directions. Combining this with Eqs.~(\ref{eq3.10}) and (\ref{eq3.13}),
one obtains
\begin{equation}
\Omega_d = \left( \frac{\widehat{\Sigma}}{d \sigma_0}
                                    \right)^d \; ,
                         \label{eq3.14}
\end{equation}
so that the radius of the critical droplet from Eq.~(\ref{eq3.11}) is
given by the standard relation
\begin{equation}
R_{\mathrm{c}} = \frac{(d-1) \sigma_0}{\Delta f} \; ,  \label{eq3.15}
\end{equation}
which, using $\Delta f$$\approx$$|H| \Delta m$, can also be written
as
\begin{equation}
R_{\mathrm{c}} = \frac{(d-1) \sigma_0}{|H| \Delta m} \; . \label{eq3.16}
\end{equation}
The size of a critical droplet therefore increases with decreasing
$|H|$, and its shape changes from a square at low temperatures to a
circle near $T_{\mathrm {c}}$.


\subsection{Finite systems}
\typeout{3.2. Finite systems}
\label{subsec3.2}
For $N^{d-1}$$\times$$\infty$ hypercylindrical systems of finite
transverse extent $N$, several distinct field regions can be
distinguished: an intermediate-field region, where
$R_{\mathrm{c}}$$\ll$$N$; a small-field region, where
$R_{\mathrm{c}}$$\stackrel{>}{_\sim}$$N$; and $H$$=$$0$, where both
phases coexist. (We do not here discuss the strong-field region,
where droplet interactions must be taken into account.)

For $R_{\mathrm{c}}$$\ll$$N$, finite-size effects are negligible,
so the decay rate of the metastable phase does not depend on $N$. All
the results derived in Sec.~\ref{subsec3.1} therefore apply in the
intermediate-field regime. For
$R_{\mathrm{c}}$$\stackrel{>}{_\sim}$$N$, on the other hand,
finite-size effects have to be taken into account. This weak-field
region coincides with the region of validity of finite-size scaling
at first-order phase transitions described in
Refs.~\cite{pri83,bor92,bor92i}.

We restrict the following discussion to $d$$=$$2$. It has recently
been shown rigorously \cite{nev91,scho92,sco93a,sco93b} that for
very low temperatures and $|H|/J$$<$$4$, a large metastable Ising
ferromagnet in two dimensions with local Metropolis dynamics decays
through a single nucleating droplet of the shape shown in
Fig.~1(a). This is an
$(l_{\mathrm{c}}$$-$$1)$$\times$$l_{\mathrm{c}}$ rectangle
of overturned spins with a single additional overturned spin as a
``knob'' on one of its long sides. The length
$l_{\mathrm{c}}$$=$$\lceil 2J/|H| \rceil$ is the smallest integer
larger than $2J/|H|$, where
$2J/|H|$$=$$\lim_{T \rightarrow 0} 2R_{\mathrm {c}}$. Flipping
the remaining $l_{\mathrm{c}}$$-$$1$ spins on the side with the
``knob'' results in a net gain of energy, and the resulting
$l_{\mathrm{c}}$$\times$$l_{\mathrm{c}}$ droplet is the
low-temperature, discrete-lattice equivalent of the critical droplet
discussed in Sec.~\ref{subsec3.1}. Finite-size effects due to the
possibility of subcritical droplets wrapping around the lattice in
the transverse direction are to be expected for
$l_{\mathrm{c}}$$\geq$$N$$-$$1$ or (extrapolating in an
approximate fashion to nonzero $T$)
$2R_{\mathrm{c}}$$>$$N$$-$$2$, which yields
$|H|$$<$$2\sigma_0/[(N$$-$$2)\Delta m]$$\leq
\! 2J/(N$$-$$2)$$\equiv$$H_2$.

\begin{figure}[ht]

\begin{center}
\begin{picture}(215,120)
\multiput(20,102.5)(15,0){10}{\line(1,0){6}}
\multiput(20,87.5)(15,0){5}{\line(1,0){6}}\put(95,85){+}
                         \multiput(110,87.5)(15,0){4}{\line(1,0){6}}
\multiput(20,72.5)(15,0){2}{\line(1,0){6}}\multiput(50,70)(15,0){5}{+}
                         \multiput(125,72.5)(15,0){3}{\line(1,0){6}}
\multiput(20,57.5)(15,0){2}{\line(1,0){6}}\multiput(50,55)(15,0){5}{+}
                         \multiput(125,57.5)(15,0){3}{\line(1,0){6}}
\multiput(20,42.5)(15,0){2}{\line(1,0){6}}\multiput(50,40)(15,0){5}{+}
                         \multiput(125,42.5)(15,0){3}{\line(1,0){6}}
\multiput(20,27.5)(15,0){2}{\line(1,0){6}}\multiput(50,25)(15,0){5}{+}
                         \multiput(125,27.5)(15,0){3}{\line(1,0){6}}
\multiput(45.5,19.5)(75.5,0){2}{\line(0,1){59}}
\put(45.5,78.5){\line(1,0){45.5}}\put(106,78.5){\line(1,0){15}}
\multiput(91,78.5)(15,0){2}{\line(0,1){15}}
\put(91,93.5){\line(1,0){15}}
\thicklines
\multiput(0,62.5)(170,0){2}{\ldots}
\multiput(0,19.5)(0,90){2}{\line(1,0){180}}
\thinlines
\put(85,0){(a)}
\put(205,59.5){$N$}
\multiput(205,19.5)(0,90){2}{\line(1,0){10}}
\put(210,19.5){\line(0,1){35}}\put(210,69.5){\line(0,1){40}}
\end{picture}

\vspace{1.0truecm}

\begin{picture}(215,120)
\multiput(20,102)(15,0){10}{\line(1,0){6}}
\multiput(20,87.5)(15,0){5}{\line(1,0){6}}\put(95,85){+}
                        \multiput(110,87.5)(15,0){4}{\line(1,0){6}}
\multiput(20,72.5)(15,0){5}{\line(1,0){6}}\put(95,70){+}
                        \multiput(110,72.5)(15,0){4}{\line(1,0){6}}
\multiput(20,57.5)(15,0){5}{\line(1,0){6}}\put(95,55){+}
                        \multiput(110,57.5)(15,0){4}{\line(1,0){6}}
\multiput(20,42.5)(15,0){5}{\line(1,0){6}}\put(95,40){+}
                        \multiput(110,42.5)(15,0){4}{\line(1,0){6}}
\multiput(20,27.5)(15,0){5}{\line(1,0){6}}\put(95,25){+}
                        \multiput(110,27.5)(15,0){4}{\line(1,0){6}}
\multiput(91,19.5)(15,0){2}{\line(0,1){74}}
\put(91,93.5){\line(1,0){15}}
\thicklines
\multiput(0,62.5)(170,0){2}{\ldots}
\multiput(0,19.5)(0,90){2}{\line(1,0){180}}
\thinlines
\put(85,0){(b)}
\put(205,59.5){$N$}
\multiput(205,19.5)(0,90){2}{\line(1,0){10}}
\put(210,19.5){\line(0,1){35}}\put(210,69.5){\line(0,1){39.5}}
\end{picture}
\end{center}

\caption[]{Sketch of two critical excitations that contribute to the
decay of the metastable phase at weak positive fields and low
temperatures. Configurations shaped like the one depicted in (a)
dominate for $|H|$$\stackrel{>}{_\sim}$$H_2$$=$$2J/(N-2)$. The
``rod''-like configuration shown in (b) prevails for
$|H|$$\stackrel{<}{_\sim}$$H_1$$=$$2J/(N-1)$. See details in
Sec.~\protect\ref{subsec3.2}.}
\label{fig1}
\end{figure}
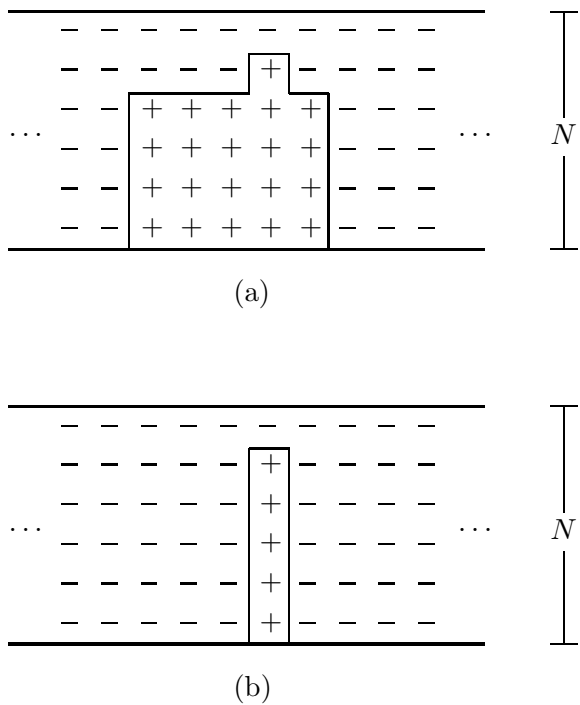

However, critical droplets of equilibrium shape are not the only
configurations comparable to the system size that may nucleate the
stable phase by wrapping around the lattice. It is easy to show that
for $|H|$$<$$H_2$, still lower-energy configurations that are
capable of spanning the system by flipping only one single spin can
be constructed by successively removing one layer of $+$ spins from
one of the short sides of the droplet shown in Fig.~1(a),
until one reaches the $(N$$-$$1)$$\times$$1$ ``rod''-shaped cluster
shown in Fig.~1(b). For
$H_2$$>$$|H|$$>$$\! 2J/(N$$-$$1)$$\equiv$$H_1$, a further energy
reduction can be achieved by adding $(N$$-$$1)$$\times$$1$ slices
back onto one of the long sides of the cluster in Fig.~1(b),
until one reaches an $(N$$-$$1)$$\times$$(N$$-$$1)$ cluster. This
does not hold true for $|H|$$<$$H_1$. Here the ``rod''-shaped cluster
of Fig.~1(b) is the lowest-energy configuration that can
nucleate the stable phase by flipping a single spin.

The above observations lead to the following predictions for the
finite-size scaling of $|\mathrm{Im} f_{\mathrm{ms}}|$. For
$|H|$$>$$H_2$, finite-size effects should be negligible, for
$H_1$$<$$|H|$$<$$H_2$ the behavior should be rather complicated
due to the large number of competing clusters with nearly degenerate
energies, and for $|H|$$\stackrel{<}{_\sim}$$H_1$ the behavior should
be determined by ``rod''-shaped clusters like the one shown in
Fig.~1(b).

To leading order, the free-energy cost $F_{\mathrm{c}}$ of creating
the ``rod''-shaped critical cluster in Fig.~1(b) is given by
its surface free energy only, so that
$F_{\mathrm{c}}$$\approx$$2 \sigma(T,H) N$. In analogy with the
equilibrium surface tension, $\sigma(T,H)$ here denotes the ``surface
tension'' of the critical droplet at non-zero fields. Inserting
$F_{\mathrm{c}}$ into Eq.~(\ref{eq3.7}), one obtains
\begin{equation}
|\mathrm{Im}f_{\mathrm{ms}}| \propto \mathrm{e}^{-2 \beta
                             \sigma(T,H) N} \; .      \label{eq3.17}
\end{equation}
The largest corrections to Eq.~(\ref{eq3.17}) should include a bulk
term in the exponential, which is of the form $-2|H|(N-1)$ for the
particular critical excitation considered here, and an $N$-dependent
prefactor.

In the infinite-system limit $H_2$$\rightarrow$$0$. We then
expect that $\sigma(T,H) \rightarrow$$\sigma_{\mathrm{eq}}(T)$ for
large $N$, where $\sigma_{\mathrm{eq}}(T)$ is the exact equilibrium
surface tension as obtained from Onsager's solution of the
two-dimensional Ising model \cite{zia82,avr82,pri83}.

At $H$$=$$0$ one can show (see Appendix \ref{appa}) that
$|\mathrm{Im}f_{\mathrm{ms}}|$ as defined in the CTM formalism,
Eq.~(\ref{eq2.20}), is given by
\begin{equation}
|\mathrm{Im}f_{\mathrm{ms}}| \propto (N \xi_N)^{-1} \; . \label{eq3.18}
\end{equation}
Here, $\xi_N$$\equiv$$\ln|\lambda_0/\lambda_1|$ denotes the
correlation length which at $H$$=$$0$ is \cite{pri83,bor92,bor92i}
\begin{equation}
\xi_N \propto N^{1/2} \mathrm{e}^{\beta \sigma_{\mathrm{eq}}(T) N} \; ,
                                                       \label{eq3.19}
\end{equation}
with a nonuniversal proportionality constant \cite{bor92}. Inserting
Eq.~(\ref{eq3.19}) into \linebreak
Eq.~(\ref{eq3.18}) one obtains
\begin{equation}
|\mathrm{Im}f_{\mathrm{ms}}| \propto N^{-3/2} \mathrm{e}^{-\beta
                      \sigma_{\mathrm{eq}}(T) N} \; .  \label{eq3.20}
\end{equation}
Note that the coefficient of $\sigma_{\mathrm{eq}}$ in
Eqs.~(\ref{eq3.17}) and (\ref{eq3.20}) differ by a factor of two.
Physically, this corresponds to the fact that, whereas the rod-like
droplets that dominate for $|H|$$\stackrel{<}{_\sim}$$H_1$ correspond
to pairs of tightly bound transverse interfaces, the interfaces are
unbound for $H$$=$$0$ \cite{bor92}.


\section{Numerical Results and Discussion}
\typeout{4. Numerical Results and Discussion}
\label{sec4}
Since the metastable phase has the same symmetry as the stable
phase \cite{mcc78,pri82a,pri82b}, which is symmetric under cyclic
permutations and reflections, we consider only those eigenstates
of the TM that are symmetric under cyclic permutations and
reflections. As an example, in Fig.~2 we plotted the
eigenvalue spectrum as $-\ln\lambda_\alpha/(\beta N)$
(Fig.~2(a)), which corresponds to the first term in
Eq.~(\ref{eq2.21}), and the constrained magnetizations
(Fig.~2(b)) for
$T/J$$=$$1.0$ ($T/T_{\mathrm{c}}$$=$$0.44068 \ldots$) and
$N$$=$$8$. When the eigenvalue spectrum is plotted as in
Fig.~2(a), the branch corresponding to $\alpha$$=$$0$ is
the equilibrium free-energy density, Eq.~(\ref{eq2.6}). In
Fig.~2(b), it corresponds to the branch with magnetization
close to $+1.0$.

As seen in Fig.~2(a), for weak magnetic fields, certain
eigenvalues group together into fans. In order to identify which
eigenvalues correspond to which subsystem configurations, we
calculated for $T$$=$$0$ the internal energies, $U_\alpha$,
and the magnetization, $M_\alpha$, for all those subsystem
configurations that are symmetric under cyclic permutations and
reflections. Configurations with the same number of interfaces
parallel to the transfer direction have the same value of $U_\alpha$.
We also find that at $H$$=$$0$ and $T$$>$$0$, the averages
calculated over all those $-\ln\lambda_\alpha/(\beta N)$ that belong
to the same fan are nearly equal to the $T$$=$$0$ values of
$U_\alpha$. (For example for $N$$=$$10$ and $T/J$$=$$1.0$,
the discrepancies between $U_\alpha$ and the corresponding averages
are smaller than $2\%$.) We therefore conclude that all the
eigenvalues that belong to the same fan correspond to subsystem
configurations with the same number of interfaces. The absence of
exact degeneracies between the eigenvalues that belong to the same
fan for $T$$>$$0$ is an entropy effect. Finally, we find that
away from the crossings, the negative of the slopes of the
$-\ln\lambda_\alpha/(\beta N)$ are approximately given by the
magnetizations $M_\alpha$ (with discrepancies that are again smaller
than $2\%$).

In accordance with earlier transfer-matrix studies
\cite{new77,mcc78,pri82a,pri82b,new80,rik92a,gor94a,fii94,gor94b,nov86},
the eigenstate $|\alpha \rangle$ corresponding to the metastable
phase at a given field was identified as the one with the largest
magnetization opposite in direction to the magnetic field (marked by
a thick solid line in Fig.~2 in the field region where this
$-\ln \lambda_\alpha$ is not the uppermost branch). As was already
observed previously \cite{mcc78}, such an eigenstate can be
identified. Furthermore, the metastable eigenvalues vary little with
$N$, indicating the existence of a well-defined thermodynamic limit
for that state \cite{mcc78}.

At certain values of the field the metastable branch seems to
cross different eigenvalue branches. These are in fact avoided
crossings, which only become exact crossings at $T$$=$$0$.
Consequently, the metastable branch does not consist of only one
eigenstate, but is composed of a succession  of eigenstates at
different fields. For $T$$=$$0$, we can use the relation
$-\ln\lambda_\alpha/(\beta N)$$=$$U_\alpha$$-$$M_\alpha H$
to calculate exactly the value of the magnetic field at which the
$m^{\mathrm{th}}$ crossing involving the metastable branch occurs:
\begin{equation}
 H_0 = 0, \mbox{ and } H_m = \frac{2J}{N-m},
   \mbox{ \ \ \ \ with \ $m$$=$$1, \ldots, N$$-$$1$} \; .
                            \label{eq4.1}
\end{equation}
For the non-zero temperatures studied here, the field values of the
avoided crossings involving the metastable branch deviate by at most
$3\%$ from their values at $T$$=$$0$. For $m$$=$$1$ and
$m$$=$$2$ Eq.~(\ref{eq4.1}) corresponds to the definitions of
$H_1$ and $H_2$ given in Sec.~\ref{subsec3.2}.

 From Eq.~(\ref{eq4.1}), the number of avoided crossings in any
neighborhood of $H$$=$$0$ is seen to go to infinity as
$N$$\rightarrow$$\infty$. This is consistent with the existence
of an essential singularity in the free energy at $H$$=$$0$
in the thermodynamic limit
\cite{lan67,lan68,lan69,fis67,gue80,zia81,low80,bak80,bax79,ent80,har84,ham83,jac83,per84,isa84,abr92,abr92i,abr93}.
In particular, it is reminiscent of the results of recent studies
which found an essential singularity in the susceptibility at
$H$$=$$0$ \cite{abr92,abr92i,abr93}. In that work the
susceptibility was calculated in an ensemble in which the
magnetization was restricted to negative values for positive
fields. It was found that in the limit $H$$\rightarrow$$0$,
the singularity manifests itself through an infinite number of
poles on the positive real axis, whose locations are given by
$H$$\propto$$1/n$, $n$$=$$1,2, \ldots, \infty$.
However, there is specific disagreement between the values of the
fields at which these poles occur and those corresponding to the
avoided crossings in the TM spectrum. Since the TM eigenvalue
crossings coincide with singularities in the metastable lifetimes
recently found in rigorous low-temperature calculations \cite{nev91},
we believe the numerical discrepancy with
Refs.~\cite{abr92,abr92i,abr93} may be due to approximations used
in that work.

Figure 3 shows the real parts (Fig.~3(a)) and the
imaginary parts (Fig.~3(b)) of the constrained free-energy
densities for $T/J$$=$$1.0$ and $N$$=$$8$. The composite
branch corresponding to the metastable phase is marked by thick
solid lines in both panels. The general features of the stable and
metastable phases in Fig.~2(a) are reproduced in
Fig.~3(a). For the real part of $f_\alpha$ the composite
metastable branch is, except near the crossings, almost identical
to the composite metastable branch in the eigenvalue spectrum,
shown in Fig.~2(a). In the spectrum of the imaginary parts,
the composite branch corresponding to the metastable phase consists
of different lobes, each of which corresponds to a different
eigenstate. The crossings for the metastable $|\mathrm{Im} f_\alpha|$
correspond to avoided crossings in the eigenvalue spectrum of
Fig.~2(a). Split lobes (such as the one at
$H/J$$\approx$$0.8$) occur when the metastable branch intersects
branches from fans corresponding to more than two interfaces. Notice
the extreme smallness of the values of the minima of the metastable
$|\mathrm{Im}f_{\alpha}|$, especially for weak magnetic fields, and
how their range extends over roughly ten decades. We also observe, in
accordance with our expectations for the finite-size scaling behavior
of $|\mathrm{Im}f_{\mathrm{ms}}|$ outlined in Sec.~\ref{subsec3.2},
that the value of the metastable $|\mathrm{Im}f_{\alpha}|$ at the
first minimum is approximately given by the square of its value at
$H$$=$$0$.

In order to avoid complications introduced by the mixing of two or
more eigenvectors in the vicinity of the crossings (where, for
example, the constrained magnetizations of the metastable eigenstates
deviate appreciably from $-1$), we concentrate on the minima of
$|\mathrm{Im}f_{\alpha}|$, which are located away from the crossings.
We denote the values of the fields at which these minima occur
by $H_m^{\mathrm{min}}$, such that
$H_{m-1}$$<$$H_m^{\mathrm{min}}$$<$$H_m$ for $m$$=$$1,
\ldots N$$-$$1$ where $H_m$ are the crossing fields of
Eq.~(\ref{eq4.1}).

Figure 4 is a semi-log plot of only the metastable
$|\mathrm{Im}f_{\alpha}|$ vs.\ inverse field, $J/H$, for
$N$$=$$9$ and $N$$=$$10$ at $T/J$$=$$1.0$. The thick
straight line was drawn through the two minima for $N$$=$$10$
between $J/H$$=$$3.0$ and $J/H$$=$$4.0$, but, in accordance
with Eq.~(\ref{eq3.7a}), it also connects minima at smaller values of
$J/H$ for both $N$$=$$9$ and $N$$=$$10$. In agreement with
our expectations outlined in Sec.~\ref{subsec3.2} above, we find
entirely different dependences on $N$ and $H$ for the metastable
$|\mathrm{Im}f_{\alpha}|$ in the weak-field and the
intermediate-field regions. As is apparent from Fig.~4,
for $H$$\geq$$H_3^{\mathrm{min}}$ $|\mathrm{Im}f_{\alpha}|$ is
independent of $N$. For $H$$\leq$$H_2^{\mathrm{min}}$, however,
$|\mathrm{Im}f_{\alpha}|$ depends strongly on $N$.
\begin{table}
\caption[]{The diameters of critical droplets, $2R_{\mathrm{c}}(T,H)$,
evaluated at (a) $H_1^{\mathrm{min}}(N)$, (b) $H_2^{\mathrm{min}}(N)$,
and (c) $H_{N-1}^{\mathrm{min}}(N)$. Here $N$ is the largest system
size used at a particular field and temperature. At
$H$$=$$H_{N-1}^{\mathrm{min}}(N)$ and for $T/J$$\geq$$1.1$,
the minimum of the metastable $|\mathrm{Im}f_\alpha|$ was so shallow
that we were unable to locate it.}
\renewcommand{\arraystretch}{.8}
\label{tab1}
\begin{center}
\begin{tabular}{lcccc}
\hline\hline\cline{2-5}\rule[-4pt]{0pt}{16pt}
(a)\ \ \ \ \ & $T/J$ & $H_1^{\mathrm{min}}(N)/J$ &
$2R_{\mathrm{c}}(T,H_1^{\mathrm{min}}(N))$ & $N$ \\
\cline{2-5}
\rule{0pt}{12pt}     &  0.4  &       0.330       &      6.0     &  6  \\
                     &  0.6  &       0.141       &     13.9     &  9  \\
                     &  0.8  &       0.084       &     22.3     & 10  \\
                     &  0.9  &       0.072       &     25.1     & 10  \\
                     &  1.0  &       0.064       &     27.1     & 10  \\
                     &  1.1  &       0.058       &     28.3     & 10  \\
                     &  1.2  &       0.054       &     28.6     & 10  \\
\cline{2-5}
\hline\cline{2-5}\rule[-4pt]{0pt}{16pt}
(b)\ \ \ \ \ & $T/J$ & $H_2^{\mathrm{min}}(N)/J$ &
$2R_{\mathrm{c}}(T,H_2^{\mathrm{min}}(N))$ & $N$ \\
\cline{2-5}
\rule{0pt}{12pt}     &  0.4  &       0.309       &     6.5      &  8  \\
                     &  0.6  &       0.234       &     8.4      & 10  \\
                     &  0.8  &       0.231       &     8.1      & 10  \\
                     &  0.9  &       0.231       &     7.8      & 10  \\
                     &  1.0  &       0.228       &     7.6      & 10  \\
                     &  1.1  &       0.224       &     7.3      & 10  \\
                     &  1.2  &       0.224       &     7.0      & 10  \\
\cline{2-5}
\hline\cline{2-5}\rule[-4pt]{0pt}{16pt}
(c)\ \ \ \ \ & $T/J$ & $H_{N-1}^{\mathrm{min}}(N)/J$ &
$2R_{\mathrm{c}}(T,H_{N-1}^{\mathrm{min}}(N))$  & $N$ \\
\cline{2-5}
\rule{0pt}{12pt}     &  0.4  &       1.838       &     1.1      &  9  \\
                     &  0.6  &       1.764       &     1.1      & 10  \\
                     &  0.8  &       1.723       &     1.1      & 10  \\
                     &  0.9  &       1.732       &     1.0      & 10  \\
                     &  1.0  &       1.781       &     1.0      & 10  \\
\cline{2-5}
\end{tabular}
\end{center}
\end{table}
This is further supported by Table~\ref{tab1}, where we compare the
diameter, $2R_{\mathrm{c}}$, of the critical droplet with the system
size, $N$, at three different values of the field,
$H_1^{\mathrm{min}}$, $H_2^{\mathrm{min}}$, and
$H_{N-1}^{\mathrm{min}}$. We find that for all temperatures studied,
the size of the critical droplet exceeds $N$ at $H_1^{\mathrm{min}}$,
is comparable to $N$ at $H_2^{\mathrm{min}}$, and is close to unity
at $H_{N-1}^{\mathrm{min}}$. The quantity $R_{\mathrm{c}}$ was
calculated from Eq.~(\ref{eq3.16}) using for $\sigma_0$ the exact
equilibrium surface tension, $\sigma_{\mathrm{eq}}(T)$, as obtained
from Onsager's solution of the two-dimensional Ising model
\cite{zia82,avr82,pri83}. We also set
$\Delta f$$=$$2m_{{\mathrm{eq}}}(T)H$, where
\begin{equation}
m_{{\mathrm{eq}}}(T) = (1-\mathrm{cosech}^{4} 2 \beta J)^{1/8}
                                                       \label{eq4.2}
\end{equation}
is the exact zero-field equilibrium magnetization \cite{yan52}.
These identifications will be justified below.

The different dependences of the metastable $|\mathrm{Im}f_{\alpha}|$
on $N$ and $H$ are discussed in more detail in the two subsections
below.


\subsection{The weak-field region}
\typeout{4.1. The weak-field region}
\label{subsec4.1}
Figure 5 shows a semi-log plot of the metastable
$N|\mathrm{Im}f_\alpha|$ (plotted as {\Large $\ast$}) and
$\xi_N^{-1}$ (shown as {\Large {\bf $\circ$}}) vs.\ $N$ at $H$$=$$0$
and $T/J$$=$$1.0$. The dashed straight lines are guides to the eye
drawn through the points at $N$$=$$9$ and $N$$=$$10$. The fact that
these lines also approximately connect points at smaller values of
$N$ indicates that, in accordance with Eqs.~(\ref{eq3.19}) and
(\ref{eq3.20}), the metastable $|\mathrm{Imf}_\alpha|$ and
$\xi_N^{-1}$ vanish exponentially with $N$.

The surface tension was calculated from either Eq.~(\ref{eq3.19})
or Eq.~(\ref{eq3.20}) by taking either the difference of the
logarithms of the metastable $|\mathrm{Im}f_\alpha|$ between two
successive $N$ or the difference of the logarithms of $\xi_N^{-1}$
between two successive $N$. Taking into account appropriate
correction terms arising from the $N$-dependent prefactors led to
estimates for the surface tension that agree very well with
$\sigma_{\mathrm{eq}}(T)$, for both the metastable
$|\mathrm{Im}f_\alpha|$ and $\xi_N^{-1}$. Except for the lowest
temperature studied (which is hampered by numerical underflow), the
differences between our estimates and $\sigma_{\mathrm{eq}}(T)$ are
smaller than $2\%$, for both the metastable $|\mathrm{Im}f_\alpha|$
and $\xi_N^{-1}$.

For the weak-field region, Fig.~6 shows a semi-log plot of
the metastable $|\mathrm{Im}f_\alpha|$ at $H_1^{\mathrm{min}}$
vs.\ $N$ for all the temperatures studied. In accordance with
Eq.~(\ref{eq3.17}), we find that the metastable
$|\mathrm{Im}f_\alpha|$ vanishes exponentially with $N$. The slopes
of the straight lines in Fig.~6, divided by $-2 \beta$,
give an estimate for $\sigma(T,H)$ at the different temperatures.

We calculated $\sigma(T,H)$ by taking the difference of the logarithms
of the metastable $|\mathrm{Im}f_\alpha|$ in Eq.~(\ref{eq3.17}) at
two different system sizes. In order to eliminate noticeable effects
due to even-odd or odd-even pairs of system sizes, $\sigma(T,H)$ was
obtained by using only odd-odd or even-even pairs of $N$. According
to Eq.~(\ref{eq4.1}), the value of the field, $H_1$, where the first
crossing occurs, converges to zero in the limit
$N$$\rightarrow$$\infty$. Therefore,
$H_1^{\mathrm{min}}$$<$$H_1$ also converges to zero, so that
in the infinite-system limit, $\sigma(T,H)$ should converge to the
equilibrium surface tension $\sigma_{\mathrm{eq}}(T)$. Assuming
linear convergence, three different values of $\sigma(T,H)$, as
obtained from Eq.~(\ref{eq3.17}), are required in order to
extrapolate $\sigma(T,H)$ to the infinite-system limit. Again, to
avoid effects due to even-odd or odd-even combinations of system
sizes, only the $\sigma(T,H)$ calculated from system sizes with
the same parity were used to extrapolate $\sigma(T,H)$ to the
infinite-system limit. The resulting extrapolations were then
averaged to obtain an estimate of $\sigma(T)$ in the infinite-system
limit. This procedure could not be carried out at $T/J$$=$$0.4$
due to numerical underflow for $N$$\geq$$6$. Instead, the value
of $\sigma(T)$ quoted at this temperature corresponds to $\sigma(T)$
as calculated directly from Eq.~(\ref{eq3.17}) using $N$$=$$4$
and $6$.

In Fig.~7 we compare $\sigma(T)$ obtained from
Eq.~(\ref{eq3.17}) (shown as {\bf $\times$}) with the exact
equilibrium surface tension, $\sigma_{\mathrm{eq}}(T)$ (shown as the
solid curve), for all the temperatures studied. Except for the lowest
temperature, the agreement between the extrapolated result and
$\sigma_{\mathrm{eq}}(T)$ is excellent, indicating that our
assumptions about the significance of $|\mathrm{Im}f_\alpha|$ and
about the shape of the critical excitation in the weak-field region
are correct. It also suggests that in the thermodynamic limit the
surface free energy of the critical excitation is given by the exact
equilibrium surface free energy.

We tried to improve the estimates of $\sigma(T)$ by including various
corrections. As an example, in Fig.~7 we also plot
extrapolated results obtained from incorporating a bulk term of the
form $-2H(N-1)$ in the exponential in Eq.~(\ref{eq3.17}) (shown as
{\bf $\Box$}). The fact that we do not find universal improvement
over the results obtained without corrections indicates that other
terms besides a bulk term might be equally important. However,
including various $N$- or $H$-dependent prefactors into
Eq.~(\ref{eq3.17}) (as in Eq.~(\ref{eq3.19}) or Eq.~(\ref{eq3.20}))
separately or in conjunction with the bulk term did not consistently
improve our previous results. These findings, together with the small
difference between all our estimates, imply that the surface term in
the free energy of the critical excitation is the dominant one.

Figure 8(a), where $H_1^{\mathrm{min}}$ is plotted vs.~$N$
for various temperatures, shows that $H_1^{\mathrm{min}}$ scales as
\begin{equation}
H_1^{\mathrm{min}}/J = c_1(T) N^{-\alpha(T)} \; .     \label{eq4.3}
\end{equation}
Here $c_1(T)$ is a proportionality factor. The temperature dependence
of the exponent $\alpha(T)$ is shown in Fig.~8(b). It is
approximately two for $T/J$$=$$1.0$ and $1.2$ and drops to a
value below $1.5$ for $T/J$$=$$0.4$. Since
$H_1^{\mathrm{min}}$$\leq$$H_1$, according to Eq.~(\ref{eq4.1}),
$\alpha(T)$$\geq$$1$ for all temperatures. We also expect
$\lim_{T \rightarrow T_{\mathrm{c}}} \alpha(T)$$=$$2$, corresponding
to a correlation length that grows linearly with $N$ at the critical
temperature, in agreement with critical finite-size scaling theory
\cite{nig90}. However, neither the origin of the $N$ dependence of
$H_1^{\mathrm{min}}$, nor the detailed temperature dependence of the
exponent $\alpha(T)$, is apparent to us at this time.


\subsection{The intermediate-field region}
\typeout{4.2. The intermediate-field region}
\label{subsec4.2}
A very different scaling behavior is found for
$H_2^{\mathrm{min}}$$\stackrel{<}{_\sim}$$H$$\stackrel{<}{_\sim}$$2J$.
In this field interval, Fig.~4 indicates that the
logarithm of the metastable $|\mathrm{Im}f_{\alpha}|$ is independent
of $N$. Only at fields $H$$\approx$$H_2^{\mathrm{min}}$ do we
find some remaining finite-size effects, which are
$\stackrel{<}{_\sim}$$3\%$. At stronger fields they vanish almost
completely, becoming smaller than $0.1\%$. If, as we propose, the
metastable $|\mathrm{Im}f_{\alpha}|$ is related to the free energy of
a critical droplet by Eq.~(\ref{eq3.7}), then this field interval of
$N$-independent $|\mathrm{Im}f_{\alpha}|$ should correspond to
critical droplet sizes between $N$ and unity. This was already seen
in Tables~\ref{tab1}(b) and \ref{tab1}(c). For all temperatures
studied, we found that the diameter of the average critical droplet
ranges from about eight lattice units for $N$$=$$10$ near the
crossover field $H_2^{\mathrm{min}}$, down to one lattice unit for
all $N$ near $H/J$$=$$2$.

It can already be inferred from the thick straight line in
Fig.~4 that the leading-order field dependence of the
metastable $|\mathrm{Im}f_{\alpha}|$ at intermediate fields can be
described by Eq.~(\ref{eq3.7a}) with an approximately
field-independent value for the parameter $\Xi$. The slope of this
line gives a rough estimate for $\Xi$, which was defined in
Eq.~(\ref{eq3.5}). Below, we emphasize the lack of an $H$ dependence
in $\Xi$, $\widehat{\Sigma}$, $\sigma_0$, and $m_{\mathrm{eq}}$ by
using the explicit notation $\Xi(T)$, $\widehat{\Sigma}(T)$,
$\sigma_0(T)$, and $m_{\mathrm{eq}}(T)$.

In order to determine $\Xi(T)$, $b$, and $B(T)$ in a more systematic
way, and thereby to study the validity of Eq.~(\ref{eq3.7a}) for
intermediate fields in greater detail, we performed linear
least-squares fits on the logarithm of the minima of the metastable
$|\mathrm{Im}f_{\alpha}|$, excluding terms of order higher than two
in the exponent. That leaves four independent parameters. However,
assuming that $B(T)$ is independent of the field, it can be
eliminated by differentiating $\ln|\mathrm{Im}f_{\mathrm{ms}}|$
with respect to the inverse magnetic field. With
$|\mathrm{Im}f_{\mathrm{ms}}|$ as given in Eq.~(\ref{eq3.7a}) and
$d$$=$$2$, one obtains
\begin{equation}
\frac{\mathrm{d} \ln|\mathrm{Im}f_{\mathrm{ms}}|}{\mathrm{d}(1/|H|)} =
                -\left[ \beta \Xi(T) + b |H| + O(H^2) \right] \; .
                                   \label{eq4.4}
\end{equation}
In order to utilize Eq.~(\ref{eq4.4}), we use the two-point
finite-difference estimate for the derivative of
$\ln|\mathrm{Im}f_{\mathrm{ms}}|$ based on the metastable
$|\mathrm{Im}f_\alpha|$ at two successive minima,
$H_m^{\mathrm{min}}$ and $H_{m+1}^{\mathrm{min}}$:
\begin{eqnarray}
\frac{\Delta(\ln|\mathrm{Im}f_\alpha|)}{\Delta (1/|H|)} & \equiv &
      \frac{\ln|\mathrm{Im}f_\alpha(H_m^{\mathrm{min}})| -
            \ln|\mathrm{Im}f_\alpha(H_{m+1}^{\mathrm{min}})|}
           {1/|H_m^{\mathrm{min}}| - 1/|H_{m+1}^{\mathrm{min}}|}
                                         \nonumber  \\
& = & - \left[ \beta \Xi(T) + b H_{\mathrm{eff}} + O(H^2) \right] \; ,
                                                 \label{eq4.5}
\end{eqnarray}
with
\begin{equation}
H_{\mathrm{eff}} \equiv \left| \frac{H_m^{\mathrm{min}}
                                     H_{m+1}^{\mathrm{min}}}
                                    {H_m^{\mathrm{min}} -
                                     H_{m+1}^{\mathrm{min}}} \right|
                    \ln \left| \frac{H_m^{\mathrm{min}}}
                                    {H_{m+1}^{\mathrm{min}}}
                                                             \right|
                                         \; .    \label{eq4.6}
\end{equation}
We then perform three-parameter fits on
$\Delta(\ln[\mathrm{Im}f_\alpha])/\Delta (1/|H|)$.

In the following, all the results reported for $\Xi(T)$ and $b$
were obtained from three-parameter fits, whereas the results given
for $B(T)$ were calculated from two-parameter fits with $\Xi(T)$ and
$b$ fixed. It should be emphasized, however, that since our results
for $|\mathrm{Im}f_{\alpha}|$ are numerically exact, we are using
the least-squares method merely as a tool to determine the above
parameters numerically.  We have therefore assigned error bars such
that $\chi^2$ per degree of freedom is approximately unity.

Typical results of the three-parameter linear least-squares fits
are shown in Fig.~9. The ratio
$-\Delta(\ln|\mathrm{Im}f_\alpha|)/\Delta (1/|H|)$
vs.~$H_{\mathrm{eff}}$ is plotted at four different temperatures.
The symbols represent the results obtained with the CTM method,
except at zero field, where we plotted the numerically exact
$\beta \Xi_{{\mathrm{eq}}}(T)$$\equiv$$\beta
\widehat{\Sigma}_{\mathrm{eq}}^2(T)/(4 \Delta m)$, with
$\Delta m$$\equiv$$2 m_{\mathrm{eq}}(T)$. Here
$\widehat{\Sigma}_{\mathrm{eq}}(T)$ was obtained by integrating
numerically over the equilibrium droplet shape, which in turn was
calculated by combining the Wulff construction with Onsager's exact
zero-field anisotropic surface tension \cite{zia82,avr82}. The
equilibrium magnetization $m_{\mathrm{eq}}(T)$ is given by
Eq.~(\ref{eq4.2}) above \cite{yan52}. The solid curves represent our
fits. The dashed straight lines start at the exact
$\beta \Xi_{{\mathrm{eq}}}(T)$ at zero field and have a slope of
unity, corresponding to the value of $b$$=$$1$, expected from field
theory, Eq.~(\ref{eq3.8}). It should be clear from Fig.~9
that our data in conjunction with the exact result at zero field
can not be fitted by straight lines which would only involve $b$
and $\Xi(T)$ as free parameters. The reason lies in the competition,
in Eq.~(\ref{eq3.7a}), between the $O(H^2)$ correction term in the
exponential, which dominates for strong fields, and the prefactor,
which becomes important for weaker fields. The accessible fields are
sufficiently strong that the $O(H^2)$-term is crucial and has to be
included into the fits. Furthermore, the field at which the
domination changes from one term to the other decreases as the
temperature is lowered. That means that for lower temperatures, the
range of fields we can cover does not extend as far into the
region of weaker fields where the power-law prefactor dominates, as
it does for the higher temperatures. This makes it increasingly
difficult to obtain reliable estimates for the prefactor exponent $b$
as $T$ is lowered.  In the intermediate-temperature region of
$0.9$$\leq$$T/J$$\leq$$1.1$, however, we find excellent agreement
between the exact $\beta \Xi_{\mathrm{eq}}(T)$ at zero field and the
corresponding quantity obtained from our fits.

This agreement is further illustrated in Fig.~10. It shows
$\Xi(T)$ divided by its equilibrium value, $\Xi_{\mathrm{eq}}(T)$,
vs.~$T$, as obtained from three-parameter linear least-squares fits,
including those shown in Fig.~9. For all $T/J$$>$$0.4$
these fits were performed on data corresponding to $N$$=$$10$.
Due to numerical underflow, we were limited to a maximum system size
of $N$$=$$9$ and fields $H$$\geq$$H_3^{\mathrm{min}}$ for
$T/J$$=$$0.4$. At temperatures $T/J$$\geq$$1.1$, on the
other hand, the minimum at $H_{N-1}^{\mathrm{min}}$ becomes so
shallow that we were unable to locate it. That reduces the number
of degrees of freedom in the fits for the higher temperatures,
$T/J$$\geq$$1.1$, by one as compared to the temperatures
$0.6$$\leq$$T/J$$\leq$$1.0$, making the fits for
these temperatures slightly less reliable than the others. Two sets
of data are shown. The data set corresponding to the diamonds was
obtained from fits using all the discernible minima of the metastable
$|\mathrm{Im}f_{\alpha}|$ in the interval
$H_2^{\mathrm{min}}$$\leq$$H$$\leq$$2J$. The data set
corresponding to the crosses with error bars was obtained by
excluding the minimum at $H_2^{\mathrm{min}}$. (For clarity, the
error bars for the diamonds, which are comparable to those for the
crosses, are not shown.) This was done in order to illustrate the
influence of uncontrollable finite-size effects due to competing
critical excitations present in the metastable
$|\mathrm{Im}f_{\alpha}|$ at $H_2^{\mathrm{min}}$. The largest
finite-size effects occur at the higher temperatures
$T/J$$\geq$$1.1$. Corrections due to finite-size effects
stay within the limits of the error bars for $T/J$$=$$0.8$, $0.9$,
and $1.0$ and increase for both higher and lower temperatures. In
the following, we always exclude from the fits the minimum of the
metastable $|\mathrm{Im}f_{\alpha}|$ that occurs at
$H_2^{\mathrm{min}}$.

The fits shown in Fig.~10 indicate that our results for
$\Xi(T)$ are in excellent agreement with the hypothesis that the
surface free energy of critical droplets is equal to that of
equilibrium droplets of the same size at the same temperature, and
that $\Delta m$ equals twice the exact zero-field magnetization
$m_{\mathrm{eq}}(T)$. Except for $T/J$$=$$1.1$, we find agreement to
within $10\%$ between the estimated $\Xi(T)$ and
$\Xi_{\mathrm{eq}}(T)$.

Independent confirmation of our results has been obtained by using
Eq.~(\ref{eq3.6}) in conjunction with direct measurements of the
lifetime of the metastable phase in Monte Carlo simulations
\cite{rik94,leex,nov94,novx}.

If we set $\Xi(T)$ to its equilibrium value $\Xi_{\mathrm{eq}}(T)$,
thereby reducing the three-parameter linear least-squares fit to
a two-parameter fit, we obtain estimates for the exponent $b$, which
are plotted in Fig.~11. As in Fig.~10, all the
estimates were calculated for
$H_3^{\mathrm{min}}$$\leq$$H$$\leq$$2J$ and $N$$=$$10$,
except at $T/J$$=$$0.4$, where we used $N$$=$$9$. For
$T/J$$>$$0.8$ our results are consistent with $b$$=$$1$.
Because of the aforementioned competition between the prefactor term,
which involves the exponent $b$, and the $O(H^2)$ correction term in
the exponential of Eq.~(\ref{eq3.7a}), the estimates for $b$ become
less reliable at lower temperatures. As a result, we do not believe
the large deviations of the estimate of $b$ from unity for
$T/J$$\leq$$0.8$ to be physically meaningful. The general agreement
of our estimates with $b$$=$$1$, in agreement with Eq.~(\ref{eq3.8})
and in disagreement with Eq.~(\ref{eq3.9}), confirm the importance
of surface excitations on the critical droplet in determining the
prefactor exponent $b$. Our results are fully consistent with
previous numerical estimates \cite{low80,har84,rik94,wal80}.

Figure 12 shows a comparison of $\Xi_{\mathrm{eq}}(T)$,
indicated by the solid line, with estimates for $\Xi(T)$ obtained
from a two-parameter linear least-squares fit, in which $b$ is fixed
to $1$. The droplet shape varies between a square at $T$$=$$0$ and
a circle at $T$$=$$T_{\mathrm{c}}$ (both shown for the whole
temperature range by dashed lines). As in Fig.~10, we find
close agreement between $\Xi(T)$ for the critical droplet with the
corresponding value $\Xi_{\mathrm{eq}}(T)$ for the equilibrium
droplet. Together with Fig.~10, these results
strongly support the notion that at a given temperature, the
metastable phase decays through the formation of critical droplets
having the same shape as equilibrium droplets at the same temperature.
This also justifies the use of the exact equilibrium surface tension
to estimate the size of the critical droplets in Table~\ref{tab1}.

We also determined the prefactor $B(T)$ by reducing the
four-parameter linear least-squares fit to Eq.~(\ref{eq3.7a}) to a
two-parameter fit by setting $\Xi(T)$ to its equilibrium value and
$b$$=$$1$. Our results are displayed in Fig.~13 as
{\bf $\times$} with error bars. They are consistent with an
exponential dependence of $B(T)$ on $1/T$, shown by the dashed line
which was obtained from a weighted linear least-squares fit to our
results. This temperature dependence of $B(T)$ can be obtained by
adding an approximately temperature independent term $\Delta F$ to
the free energy of the critical droplet. Our linear least-squares fit
indicates that $\Delta F$$\approx$$1.8$. This result is supported by
a recent Monte Carlo study \cite{nov94,novx}. Using continuum droplet
theory, values of $B(T)$ were calculated from measurements of the
lifetime of the metastable phase and are shown as {\bf $\Diamond$}
with error bars in Fig.~13. They also follow a straight line
with a value of $\Delta F$$\approx$$1.2$ which is of the same order
of magnitude as the CTM result. The discrepancy between the two
estimates of $\Delta F$ is likely due to a temperature dependent
kinetic prefactor in the Monte Carlo results. Also shown in
Fig.~13 are values of $B(T)$ as calculated from the
Becker--D\"{o}ring droplet model by Harris \cite{har84}. Clearly,
the Becker--D\"{o}ring droplet model is inadequate to describe the
temperature dependence of our results. An additive term in the free
energy of the droplet was also found by Jacucci et al.\ \cite{jac83}
which they interpreted as a curvature-dependent surface free energy.
However, although their additive term is field independent, it does
depend on the temperature. Recent rigorous studies predict
$\Delta F$$=$$4J$ at very low $T$ \cite{nev91,scho92,sco93a,sco93b}.
We note that our attempts at providing a simple, heuristic
extrapolation of this result to the temperature range covered by the
CTM and Monte Carlo studies by using
$\Delta F(T)$$\approx$$2 \sigma(T)$ were not successful in fitting
the numerical results. A clarification, especially of the physical
origin of $\Delta F$ as found in the CTM and Monte Carlo results,
must await further study.

For the interpretation of the results of the CTM method in terms of
the droplet model to be consistent, the critical droplets have to
be compact with well defined boundaries. Explicit verification of this
is given in Appendix \ref{appb}.


\section{Summary and Conclusions}
\typeout{5. Summary and Conclusions}
\label{sec5}
We successfully applied the constrained-transfer-matrix method
\cite{rik89,rik91a} to the two-dimensional nearest-neighbor Ising
ferromagnet in an external magnetic field to obtain complex-valued
constrained free energies. In very good agreement with nucleation
theory, we found several different field regimes, set apart by
characteristic field and system-size dependences of the metastable
$|\mathrm{Im}f_{\alpha}|$. At $H$$=$$0$, we find that the
metastable $|\mathrm{Im}f_{\alpha}|$ is proportional to
$(N\xi_N)^{-1}$ and therefore decays exponentially with $N$. For
fields $0$$<$$H$$<$$H_1$, where the size of the critical droplet as
determined from droplet theory is larger than the transverse system
size $N$, the decay of the metastable phase proceeds through the
formation of ``rod''-like excitations that extend $(N$$-$$1)$ lattice
units in the finite direction and are only one lattice unit wide.
Introducing a ``surface tension'' for nonzero fields, this again leads
to an exponential decay of the metastable $|\mathrm{Im}f_{\alpha}|$
with $N$. Due to the binding of two interfaces in the rod-like
excitation, the argument of the exponential factor in the metastable
$|\mathrm{Im}f_{\alpha}|$ at $0$$<$$H_1^{\mathrm{min}}$$<$$H_1$ is
approximately twice as large as the one at $H$$=$$0$. Extrapolating
the nonequilibrium surface tension to the infinite system limit, we
find very good agreement with Onsager's exact equilibrium surface
tension. For $H$$>$$H_2$, the decay of the metastable phase is
dominated by critical droplets that are smaller than the finite width
of the system. The metastable $|\mathrm{Im}f_{\alpha}|$ in this field
regime is therefore $N$-independent. The functional form of the
imaginary part of the metastable free energy is the same as the one
obtained from field-theoretical droplet models for infinitely large
systems in ultraweak fields \cite{lan69,gue80}
(see Eq.~(\ref{eq3.7a})). Also, our numerical results indicate that
the critical droplets have the same surface free energy as
equilibrium droplets of the same size at the same temperature. This
and our estimates for the prefactor exponent $b$ are in excellent
agreement with series expansion \cite{low80,bak80,bax79,ent80,har84},
exact diagonalization \cite{ham83}, and Monte Carlo results
\cite{jac83,per84}. A residual $N$-dependence remains in the
transition region $H_1$$<$$H$$<$$H_2$, where a number of critical
excitations of different shapes have comparable energies. Finally, in
Appendix \ref{appb}, we have given some estimates for the field
$H_{\mathrm{MFSP}}$ beyond which standard droplet theory is expected
to become unreliable. For $H$$>$$H_{\mathrm{MFSP}}$, we could not
resolve a metastable imaginary free energy from all the other
branches of $|\mathrm{Im} f_\alpha|$.

Our results for the dependences of the metastable
$|\mathrm{Im}f_\alpha|$ on system size, field, and temperature can
be compared with direct measurements of metastable lifetimes from
Monte Carlo studies
\cite{mcc78,sto72,bin73a,bin74,sto77,ray90a,tom92,rik94,leex,nov94,novx}.
The picture that emerges from Ref.~\cite{rik94} involves four
distinct field regions, in which the decay proceeds via different
mechanisms. For very weak fields, the lifetimes obtained from the
Monte Carlo simulations grow exponentially with the area of an
interface cutting across the system, indicating that the excitations
that lead to a decay of the metastable phase are of a size comparable
to the finite dimension of the system. For intermediate fields, the
lifetime is proportional to the inverse system volume, indicating
that a small, size-independent nucleation rate is the
rate-determining step. This ``single-droplet region'' \cite{rik94} is
characterized by decay via a single critical droplet. For yet
stronger fields, the lifetime is size independent, indicating that
the rate-determining process is the nucleation and growth of a finite
density of droplets. This region was called the
``multi-droplet region'' in Ref.~\cite{rik94}. Finally, in the
``strong-field region'' beyond $H_{\mathrm{MFSP}}$, the droplet
picture becomes inadequate.

In the CTM study presented here, the ``weak-field region'', where
$|\mathrm{Im}f_\alpha|$ depends exponentially on the system size,
and the ``intermediate-field region'' correspond directly to the
weak-field and the single-droplet regions found in the Monte Carlo
study in Ref.~\cite{rik94}. In the CTM study the ``weak-field
region'' extends from $H$$=$$0$ to approximately the field of
the first eigenvalue crossing at $H_1/J$ and the ``single droplet
region'' extends from approximately $H_1/J$ to approximately
$H_{\mathrm{MFSP}}$ given in Appendix \ref{appb}, with no clearly
distinguishable ``multi-droplet region''. The values for
$H_{\mathrm{MFSP}}$ obtained in the study presented here are of the
same order of magnitude as the ones obtained by Monte Carlo
simulation \cite{rik94}.

Comparing the field and temperature dependence of
$|\mathrm{Im}f_\alpha|$ with that of the metastable lifetime from
Monte Carlo calculations \cite{rik94,leex,nov94,novx} shows that for
$H$$<$$H_{\mathrm{MFSP}}$ both quantities can be obtained to leading
order from the same droplet theory. The only difference is a kinetic
prefactor that enters into the calculation of the lifetime
\cite{lan68,lan69}, which depends on the particular dynamic chosen in
the Monte Carlo algorithm \cite{rik94}. This is in complete agreement
with and therefore supports the validity of Eq.~(\ref{eq3.7a}) not
only for ultraweak fields but also for intermediate fields.

Comparison of the results obtained for the typical
short-range-force model, presented here, with CTM studies of
two- and three-state systems with weak long-range interactions
\cite{rik89,rik91a,rik92a,gor94a,fii94} clearly shows the
differences and similarities in the properties of the metastable
phase in the two types of systems \cite{gor94b}. The lifetime in
short-range-force models becomes large but stays finite even in
the thermodynamic limit, with an imaginary part of the metastable
free energy that displays a droplet-model type essential singularity
at zero magnetic field. Long-range-force models, on the other hand,
exhibit metastable phases that become infinitely long-lived as the
range of the interactions goes to infinity \cite{pen71,pen79}, with
an imaginary part of the metastable free energy that exhibits scaling
behavior consistent with a well-defined non-zero spinodal field and
a corresponding branch-point singularity \cite{gor94a,fii94}. As the
CTM studies for those systems show, the region in which the
metastable $|\mathrm{Im}f_\alpha|$ decays exponentially with $N$
extends all the way out to the exactly known mean-field spinodal.
Nevertheless, in both types of models, the leading-order dependence
of the metastable $|\mathrm{Im}f_\alpha|$ on the magnetic field and
the temperature is, over a wide range of fields, given by a Boltzmann
factor involving a well-defined free energy of formation for a
critical excitation in the metastable phase \cite{gor94a,fii94}.

All our results strongly support the conclusion that the
constrained-transfer-matrix formalism provides a nonperturbative
method to numerically continue the equilibrium free energy into the
metastable phase that, in contrast to the analytic continuations
obtained in field-theoretic\-al droplet models, does not rely on the
explicit introduction of the particular excitations through which the
metastable phase decays.

A question of current interest is whether or not the decay of
metastability in anisotropic square-lattice Ising ferromagnets
\cite{zia85,kot93} proceeds through equilibrium-shape critical
droplets \cite{zia82,avr82,nev91}. A study of this topic by the CTM
method is in progress \cite{guexx}.


\section*{Acknowledgement}
We would like to thank R.~K.~P.\ Zia for helpful correspondence
and B.~M.~Gorman and J.~Lee for useful conversations.


\appendix
\typeout{appendix}


\section{The relationship between $|\mathrm{Im} f_{\mathrm{ms}}|$ and
         $\xi_N$ at $H$$=$$0$}
\label{appa}
As may be seen from Fig.~3, the metastable phase for
$|H|$$<$$H_1$ is represented by the eigenstate $|1 \rangle$, so that
$\mathrm{Im} f_{\mathrm{ms}}$$=$$\mathrm{Im} f_1$ at $H$$=$$0$. Using
Eqs.~(\ref{eq2.17}) through (\ref{eq2.20}) one obtains
\begin{equation}
\mathrm{Im} f_1 = -T \mathrm{Im} S_1 \; ,    \label{eqa1}
\end{equation}
where $S_1$ is given by Eq.~(\ref{eq2.19}):
\begin{equation}
S_1 = -\frac{1}{N} \sum_{x_l,x_{l+1}} P_1 (x_l,x_{l+1}) \,
  \mathrm{Ln} \langle x_l |\lambda_1^{-1} {\bf T}_1| x_{l+1}
                   \rangle \; .    \label{eqa2}
\end{equation}

With Eqs.~(\ref{eq2.12}) and (\ref{eq2.16}), it is easily shown that
\begin{equation}
\lambda_1^{-1} {\bf T}_1 = \lambda_1^{-1} {\bf T}_0 - |0 \rangle
                    \left( \frac{\lambda_0}{\lambda_1} -
                           \frac{\lambda_1}{\lambda_0} \right)
                                                     \langle 0| \; .
                                                     \label{eqa3}
\end{equation}
A term in the sum in Eq.~(\ref{eqa2}) contributes to the imaginary
part of $S_1$ only if the argument of the logarithm is negative. If
we let $Y$ denote the set of all configurations $(x_l,x_{l+1})$ such
that $\langle x_l| \lambda_1^{-1} {\bf T}_1 |x_{l+1} \rangle$$<$$0$,
then for all $(x_l,x_{l+1})$$\in$$Y$
\begin{equation}
\mathrm{Im} (\mathrm{Ln} \langle x_l |\lambda_1^{-1} {\bf T}_1| x_{l+1}
                                                    \rangle)
            = - \mathrm{i} \pi \; .   \label{eqa4}
\end{equation}
Using Eqs.~(\ref{eqa1}), (\ref{eqa2}) and (\ref{eqa4}) one obtains
\begin{equation}
\mathrm{Im} f_1 = - \pi \frac{T}{N} \sum_Y P_1(x_l,x_{l+1}) \; ,
                                                       \label{eqa5}
\end{equation}
where $\sum_Y$ denotes the sum over all configurations
$(x_l,x_{l+1})$$\in$$Y$. With Eqs.~(\ref{eq2.10}) and (\ref{eqa3})
this can be written as
\begin{eqnarray}
\mathrm{Im} f_1 = - \pi \frac{T}{N} \sum_Y \langle 1|x_l \rangle
                            \mbox{\hspace{8.8cm}}  \label{eqa6} \\
    \mbox{\hspace{2cm}} \times \left[ \langle x_l |
                         \lambda_1^{-1} {\bf T}_0|
                         x_{l+1} \rangle - \langle x_l|0 \rangle
                         \left( \frac{\lambda_0}{\lambda_1} -
                               \frac{\lambda_1}{\lambda_0} \right)
                          \langle 0|x_{l+1} \rangle
                  \right] \langle x_{l+1}|1 \rangle \; .   \nonumber
\end{eqnarray}
Using the definition of the correlation length in a system of finite
width $N$, $\xi_N^{-1}$$\equiv$$\ln|\lambda_0/\lambda_1|$, and the
fact that $\xi_N$$\rightarrow$$\infty$ for $T$$\rightarrow$$0$, one
obtains
\begin{equation}
\left( \frac{\lambda_0}{\lambda_1} -
 \frac{\lambda_1}{\lambda_0} \right) = 2 \sinh(\xi_N^{-1}) \; ,
                                               \label{eqa7}
\end{equation}
so that
\begin{eqnarray}
\mathrm{Im} f_1 = - \pi \frac{T}{N} \sum_Y \langle 1|x_l \rangle
                            \mbox{\hspace{8.8cm}}  \label{eqa8} \\
    \mbox{\hspace{2cm}} \times \left[ \langle x_l |
                         \lambda_1^{-1} {\bf T}_0|
                         x_{l+1} \rangle
                         - \langle x_l|0 \rangle 2 \sinh(\xi_N^{-1})
                           \langle 0|x_{l+1} \rangle \!
                  \right] \langle x_{l+1}|1 \rangle \; .   \nonumber
\end{eqnarray}

The quantity in square brackets in Eq.~(\ref{eqa8}) is simply
$\langle x_l |\lambda_1^{-1} {\bf T}_1| x_{l+1} \rangle$, and
therefore negative for every $(x_l,x_{l+1})$ included in the sum.
Since each of the terms in the difference inside the square brackets
is positive, we must have
\begin{equation}
0 < \langle x_l |\lambda_1^{-1} {\bf T}_0| x_{l+1} \rangle
  < \langle x_l|0 \rangle 2 \sinh(\xi_N^{-1}) \langle 0|x_{l+1}
    \rangle \; ,                \label{eqa9}
\end{equation}
and consequently, since
$\sinh(\xi_N^{-1})$$=$$\xi_N^{-1}$$+$$O(\xi_N^{-3})$, we
obtain
\begin{equation}
\mathrm{Im} f_1 = \mathrm{const} \frac{\pi T}{N}
                  \xi_N^{-1} + O(\xi_N^{-3})
             \propto (N \xi_N)^{-1} \; ,        \label{eqa10}
\end{equation}
which yields Eq.~(\ref{eq3.18}). Only two exceptions to this scenario
are possible. Exact cancelation of the leading-order terms in the
square brackets for all $(x_l,x_{l+1})$$\in$$Y$, or multiplicative
factors of $O(\xi_N^{-1})$ arising from the matrix elements with the
eigenvectors. Both effects are highly improbable and in any case
would lead to an even smaller $|\mathrm{Im}f_1|$ than that given by
Eq.~(\ref{eqa10}).

For $N$$=$$1$, ${\bf T}_1$ is a $2$$\times$$2$ matrix, so
that we can easily calculate $\mathrm{Im}f_1$ exactly, obtaining
\begin{equation}
\mathrm{Im}f_1 = -\frac{\pi T}{2} (1-\mathrm{e}^{-\xi_1^{-1}}) \; .
                                                      \label{eqa11}
\end{equation}
This gives $\mathrm{const}$$=$$-1/2$ in Eq.~(\ref{eqa10}). Under the
assumption that the full transfer matrix of the system at $H$$=$$0$
and for $N$$>$$1$ can be approximated by an effective $2$$\times$$2$
transfer matrix with eigenvalues that are identical to $\lambda_0$
and $\lambda_1$ of the full transfer matrix \cite{pri83,bor92,bor92i},
one would expect
\begin{equation}
\mathrm{Im}f_1 \approx
                  -\frac{\pi T}{2N} (1-\mathrm{e}^{-\xi_N^{-1}}) \; ,
                                                      \label{eqa12}
\end{equation}
leading to $\mathrm{const}$$\approx$$-1/2$ for $N$$>$$1$ as well. In
Fig.~A.1 we plot $2 N |\mathrm{Im}f_1|/(\pi T \xi_N^{-1})$ as
a function of $N$ for all the temperatures studied. Our expectation
is met for low temperatures and small system sizes. However, the
plotted quantity still depends rather strongly on $N$ and $T$ for
higher $T$ and $N$. This is not so surprising, since for increasing
$T$ entropy terms also increase, and for increasing $N$ the
eigenvalue spectrum becomes more complex, both effects affecting the
values of the matrix elements in the sum in Eq.~(\ref{eqa8}).

We have observed that the quantity plotted in Fig.~A.1 shows
nice data collapse onto a single scaling function when plotted
vs.~the scaling variable $\sqrt{N-1}T$. However, we have not yet
found a theoretical argument to support this empirical scaling
relation.


\section{On the applicability of the compact-droplet picture}
\label{appb}
For the interpretation of the results of the CTM method in terms of
the droplet model to be consistent, the critical droplets have to
have well defined boundaries. This requires that the size of the
critical droplets is larger than the single-phase correlation lengths
in both the stable and metastable phases. A single-phase correlation
length, $\xi_{\mathrm{sp}}$, can be defined by
\begin{equation}
\xi_{\mathrm{sp}}^{-1}=\ln \left| \frac{\lambda_\alpha}{\lambda_\beta}
                                       \right| \; .
                        \label{eqb1}
\end{equation}
For the single-phase correlation length corresponding to the stable
phase, $\lambda_\alpha$$=$$\lambda_0$ is the dominant eigenvalue
of the TM, whereas $\lambda_\beta$ is given by the eigenvalues of
the eigenstates that have magnetizations closest to $M$$=$$(N-2)/N$.
In the case of the metastable phase, $\lambda_\alpha$ are the
eigenvalues corresponding to the metastable branch in the eigenvalue
spectrum, whereas $\lambda_\beta$ are the eigenvalues corresponding
to the branches with magnetizations closest to $M$$=$$-(N-2)/N$.

An estimate for the eigenvalues $\lambda_\alpha$ can be obtained
analytically by using
\begin{equation}
\ln \lambda_\alpha \approx -\frac{N}{T}{\mathrm{Re}}f_\alpha =
     -\frac{N}{T} (U_\alpha - M_\alpha H - T {\mathrm{Re}}S_\alpha)
                                          \; ,     \label{eqb2}
\end{equation}
where $U_\alpha$, $M_\alpha$, and $S_\alpha$ are the constrained
internal energy, constrained magnetization, and constrained entropy
of the eigenstate $|\alpha \rangle$, as defined in
Eqs.~(\ref{eq2.17}), (\ref{eq2.18}), and (\ref{eq2.19}). For low
temperatures we can neglect the entropy term and approximate the
internal energy and the magnetization by their values at
$T$$=$$0$. We therefore insert $U_\alpha$$=$$-2J$,
$U_\beta$$=$$U_\alpha$$+$$(4J/N)$, and $M_\alpha$ and
$M_\beta$ as given above into Eqs.~(\ref{eqb1}) and (\ref{eqb2}).
This yields:
\begin{eqnarray}
\xi_{\mathrm{st}} & \approx & \frac{T}{2(2J+H)}   \label{eqb3}   \\
\xi_{\mathrm{ms}} & \approx & \frac{T}{2(2J-H)}   \label{eqb4} \; .
\end{eqnarray}
Here $\xi_{\mathrm{st}}$ and $\xi_{\mathrm{ms}}$ denote the
single-phase correlation length of the stable and the metastable
phase, respectively.

Figure B.1 shows a comparison between the critical diameter,
$2R_{\mathrm{c}}$, and the single-phase correlation lengths,
$\xi_{\mathrm{st}}$ and $\xi_{\mathrm{ms}}$, for $T/J$$=$$1.2$.
The solid lines were obtained from Eq.~(\ref{eq3.16}) for
$R_{\mathrm c}$ and from Eqs.~(\ref{eqb3}) and (\ref{eqb4}) for
the single-phase correlation lengths. Also shown are the single-phase
correlation lengths as obtained from TM calculations for
$N$$=$$10$. For the stable single-phase correlation length the
agreement between the TM results and Eq.~(\ref{eqb3}) is excellent.
In the case of the metastable single-phase correlation length the TM
calculations agree well with Eq.~(\ref{eqb4}) up to
$H/J$$\approx$$1$. For fields beyond $H/J$$\approx$$1$, the
increasing discrepancy between the metastable single-phase correlation
length as obtained from the TM and Eq.~(\ref{eqb4}), is caused by
entropy effects and the deviation of $U_\alpha$ and $M_\alpha$ from
their zero-temperature values.

An estimate for the magnetic field $H_{\mathrm{MFSP}}$, at which
standard continuum nucleation theory is expected to become suspect,
is given by the intersection of $\xi_{\mathrm{ms}}$ and
$2R_{\mathrm{c}}$. An estimate of the temperature dependence of
$H_{\mathrm{MFSP}}$ can be obtained by equating $\xi_{\mathrm{ms}}$
from Eq.~(\ref{eqb4}) and $2R_{\mathrm{c}}$ from Eq.~(\ref{eq3.16}).
Using the exact zero-field equilibrium surface tension
$\sigma^{\mathrm{eq}}_0(T)$ and the equilibrium magnetization
$m_{\mathrm{eq}}(T)$ we get
\begin{equation}
\frac{H_{\mathrm{MFSP}}(T)}{J} \approx
       \frac{4 \sigma^{\mathrm{eq}}_0(T)}{[2 \sigma^{\mathrm{eq}}_0(T)
                                        + T m_{\mathrm{eq}}(T)]} \; .
                                            \label{eqb5}
\end{equation}
This relation is illustrated in Fig.~B.2. As expected,
$H_{\mathrm{MFSP}}/J$$=$$2$ at $T/J$$=$$0$. However, since
Eqs.~(\ref{eqb3}) and (\ref{eqb4}) are low-temperature approximations,
Eq.~(\ref{eqb5}) must be unreliable as $T$ approaches
$T_{\mathrm{c}}$. Certainly, continuum droplet theory breaks down
when the diameter of the critical droplet becomes smaller than one
lattice unit \cite{nev91}. Using this condition one obtains
$H_{\mathrm{MFSP}}$ as shown by the dashed line in Fig.~B.2.
The $H_{\mathrm{MFSP}}$ obtained from this condition is of the same
order of magnitude as the one obtained from Eq.~(\ref{eqb5}). Also
plotted in Fig.~B.2 (shown as {\Large {\bf $\circ$}}) is
$H_{N-1}^{\mathrm{min}}(T)$ for $N$$=$$10$ (from Table~\ref{tab1}(c))
for all those temperatures that permitted us to determine it with the
CTM method. Since $H_{N-1}^{\mathrm{min}}(T)$ is the field at which
according to Table~\ref{tab1}(c) $2R_{\mathrm{c}}$$\approx$$1$, we
expect it to be of the same order of magnitude as
$H_{\mathrm{MFSP}}$. This is indeed seen to be the case. The results
summarized in this appendix confirm the applicability of the
compact-droplet picture with sharp interfaces in the entire field and
temperature range covered by our study.


\typeout{Bibliography}


\begin{figure}
\caption[]{(a) The eigenvalue spectrum,
$-\ln\lambda_\alpha/(\beta J N)$, and (b) the constrained
magnetizations $M_\alpha$ as given by Eq.~(\protect\ref{eq2.18}),
as functions of the magnetic field $H/J$ for
$T/J$$=$$1.0$ ($T/T_{\mathrm{c}}$$\approx$$0.441$) and
$N$$=$$8$. Shown are only those branches that correspond to
eigenstates that are symmetric under translation and reflection. The
segments that contribute to the metastable branch are marked by thick
lines.}
\label{fig2}
\end{figure}

\begin{figure}
\caption[]{(a) The real parts, $\mathrm{Re}f_\alpha/J$, and (b) the
imaginary parts, $|\mathrm{Im}f_\alpha|/J$, of the constrained free
energies vs.\ $H/J$, for $T/J$$=$$1.0$ and $N$$=$$8$. Notice
the overall similarity of the stable and metastable branches of
$\mathrm{Re}f_\alpha/J$ to those of the eigenvalue spectrum in
Fig.~\protect\ref{fig2}(a). The composite metastable branch is marked
by thick lines. See detailed discussion in Sec.~\protect\ref{sec4}.}
\label{fig3}
\end{figure}

\begin{figure}
\caption[]{Semi-log plot of the imaginary parts of the
constrained free energies, $|\mathrm{Im}f_\alpha|/J$, that correspond
to the metastable branch, shown vs.\ $J/H$. Plotted are data for
two different system sizes, $N$$=$$9$ and $N$$=$$10$, at
$T/J$$=$$1.0$. The thick, straight line was drawn through the two
minima between $J/H$$=$$3.0$ and $J/H$$=$$4.0$ (marked by
{\Large $\bullet$}) for $N$$=$$10$ only.}
\label{fig4}
\end{figure}

\begin{figure}
\caption[]{Comparison of the metastable $N|\mathrm{Im}f_\alpha|/J$
({\Large {\bf $\ast$}}) and $\xi_N^{-1}$ ({\Large {\bf $\circ$}})
vs.\ $N$ in a semi-log plot at $H$$=$$0$ and $T/J$$=$$1.0$. The
dashed straight lines are guides to the eye drawn through the points
at $N$$=$$9$ and $N$$=$$10$. See details in
Sec.~\protect\ref{subsec4.1}.}
\label{fig5}
\end{figure}

\begin{figure}
\caption[]{Semi-log plot of the minimum of
$|\mathrm{Im}f_\alpha|/J$ at the weak field
$0$$<$$H_1^{\mathrm{min}}$$<$$H_1$, plotted vs.\ $N$
for $T/J$$=$$0.4$, $0.6$, $0.8$, $1.0$, and $1.2$. At these weak
fields, $|\mathrm{Im}f_\alpha|$ decays exponentially with $N$ as is
evidenced by the straight lines. For each temperature, these lines
were simply drawn through the points corresponding to the two
largest system sizes available.}
\label{fig6}
\end{figure}

\begin{figure}
\caption[]{Comparison of extrapolations of the surface tension,
$\sigma(T)/J$, as obtained from Eq.~(\protect\ref{eq3.17}) (shown as
{\bf $\times$}), with the exact equilibrium surface tension,
$\sigma_{\mathrm{eq}}(T)$ (solid curve). Also shown ({\bf $\Box$})
are estimates obtained by adding a bulk term of the form $2H(N-1)$
in the exponential in Eq.~(\protect\ref{eq3.17}). See detailed
discussion in Sec.~\protect\ref{subsec4.1}.}
\label{fig7}
\end{figure}

\begin{figure}
\caption[]{(a) Log-log plot of $H_1^{\mathrm{min}}/J$
vs.\ $N$ for $T/J$$=$$0.4$, $0.6$, $0.8$, $1.0$, and $1.2$.
The straight lines show that $H_1^{\mathrm{min}}$ is proportional
to $N^{-\alpha(T)}$. (b) The exponent $\alpha(T)$ as a function of
$T$, as obtained from linear least-squares fits to our data.}
\label{fig8}
\end{figure}

\begin{figure}
\caption[]{Shown is a plot of the two-point finite-difference estimate
for $\beta \Xi(T)/J$, \linebreak
$-\Delta (\ln|\mathrm{Im}f_\alpha|)/\Delta (J/|H|)$, as a function
of $H_{\mathrm{eff}}/J$ (compare Eqs.~(\protect\ref{eq4.5}) and
(\protect\ref{eq4.6})) for four different temperatures. The symbols
represent the results obtained with the CTM method, except at zero
field, where we plotted $\beta \Xi_{\mathrm{eq}}(T)/J$ obtained from
the exact equilibrium quantities $\widehat{\Sigma}_{\mathrm{eq}}(T)$
and $\Delta m$$=$$2 m_{\mathrm{eq}}(T)$ as described in
Sec.~\protect\ref{subsec4.2}. The solid curves represent
three-parameter linear least-square fits. The dotted straight lines
have a slope of unity, corresponding to a value of
$b$$=$$1$, and they intercept the vertical axis at the exact
$\beta \Xi_{\mathrm{eq}}(T)/J$.}
\label{fig9}
\end{figure}

\begin{figure}
\caption[]{The quantity $\Xi(T)$ divided by its equilibrium value
$\Xi_{{\mathrm{eq}}}(T)$, calculated from a three-parameter linear
least-squares fit to Eq.~(\protect\ref{eq4.5}). The {\bf $\Diamond$}
correspond to results from fits including all the minima of the
metastable $|\mathrm{Im}f_\alpha|$ in the interval
$H_2^{\mathrm{min}}$$\leq$$H$$\leq$$2J$, whereas the
{\bf $\times$} with error bars were obtained by excluding
$H_2^{\mathrm{min}}$ to reduce uncontrollable finite-size effects.
(For clarity, no error bars are shown for the {\bf $\Diamond$}.)
The system size was $N$$=$$10$, except at $T/J$$=$$0.4$,
where we used $N$$=$$9$. See Sec.~\protect\ref{subsec4.2} for a
detailed discussion.}
\label{fig10}
\end{figure}

\begin{figure}
\caption[]{The prefactor exponent $b$ vs.\ $T/J$. The values shown
were obtained from a two-parameter linear least-squares fit to
Eq.~(\protect\ref{eq4.5}) by setting $\Xi(T)$ to its equilibrium
value. See Sec.~\protect\ref{subsec4.2} for details.}
\label{fig11}
\end{figure}

\begin{figure}
\caption[]{The quantity $\Xi(T)/J^2$ as obtained from a
two-parameter linear least-squares fit to Eq.~(\protect\ref{eq4.5}),
setting $b$$=$$1$. The solid line corresponds to the equilibrium
value $\Xi_{\mathrm{eq}}(T)$. The droplet shape interpolates between
a square at $T$$=$$0$, given by $(2 \sigma_0^2)/m_{\mathrm{eq}}$, and
a circle at $T$$=$$T_{\mathrm{c}}$, given by
$(\pi \sigma_0^2)/(2 m_{\mathrm{eq}})$, both of which are shown for
the whole temperature range as dashed curves. The metastable CTM
estimates, shown as $\bullet$ with error bars, follow the equilibrium
curve closely. The dashed vertical line marks the critical
temperature. See Sec.~\protect\ref{subsec4.2} for details.}
\label{fig12}
\end{figure}

\begin{figure}
\caption[]{Semi-log plot of the prefactor $B(T)$ vs.\ inverse
temperature $J/T$.
The CTM results are indicated by {\bf $\times$} with error bars
and were obtained from a two-parameter linear least-squares fit to
Eq.~(\protect\ref{eq3.7a}) by setting $\Xi(T)$ to its equilibrium
value and $b$$=$$1$. The dashed line is a weighted linear
least-squares fit of the values of $B(T)$ to the form
$B(T)$$=$$k \exp(-\Delta F/T)$, and gives
$\Delta F/J$$=$$1.78(5)$ and $k$$=$$2.12(7)$. Also shown are
Monte Carlo results for $B(T)$ ({\bf $\Diamond$} with error bars)
as obtained by continuum droplet theory from direct measurements of
the lifetime of the metastable phase \protect\cite{nov94,novx}. The
dot-dashed line is a weighted linear least-squares fit which gives
$\Delta F/J$$=$$1.25(5)$ and $k$$=$$3.54(8)$. The {\Large
{\bf $\circ$}} represent $B(T)$ as calculated from Becker--D\"{o}ring
droplet theory by Harris \protect\cite{har84}. See
Sec.~\protect\ref{subsec4.2} for details.}
\label{fig13}
\end{figure}

\renewcommand{\theequation}{\Alph{section}\arabic{equation}}
\renewcommand{\thefigure}{{\protect\thesection.\arabic{figure}}}
\setcounter{section}{0}
\setcounter{figure}{0}
\setcounter{equation}{0}

\begin{figure}
\caption[]{The quantity $2 N |{\rm Im}f_1|/(\pi T \xi_N^{-1})$ at
$H$$=$$0$ as a function of $N$ is shown for all the temperatures
studied. At $T/J$$\leq$$0.8$ numerical underflow prevented us
from obtaining reliable results for the larger $N$. An exact
calculation of $2 N |{\rm Im}f_1|/(\pi T \xi_N^{-1})$ at $H$$=$$0$
for $N$$=$$1$ leads to a value of unity for all temperatures
(shown as {\Large $\bullet$}). See Appendix \protect\ref{appa} for
more details.}
\label{figa1}
\end{figure}

\setcounter{figure}{0}
\setcounter{equation}{0}

\begin{figure}
\caption[]{Comparison of analytical estimates of the
diameter of the critical droplet, $2R_{\mathrm{c}}$, with the
estimated single-phase correlation lengths, $\xi_{\mathrm{st}}$ for
the stable phase, and $\xi_{\mathrm{ms}}$ for the metastable phase
at $T/J$$=$$1.2$. Here $2R_{\mathrm{c}}$ was calculated from
Eq.~(\protect\ref{eq3.16}), and $\xi_{\mathrm{st}}$ and
$\xi_{\mathrm{ms}}$ were obtained from Eqs.~(\protect\ref{eqb3})
and (\protect\ref{eqb4}). The transfer-matrix results for
$\xi_{\mathrm{st}}$ and $\xi_{\mathrm{ms}}$ for $N$$=$$10$ are shown
as {\bf $\Diamond$}.}
\label{figb1}
\end{figure}

\begin{figure}
\caption[]{Estimates of the crossover field $H_{\mathrm{MFSP}}/J$
obtained from Eq.~(\protect\ref{eqb5}) (solid line) and from the
condition $2R_{\mathrm{c}}$$=$$1$ (dashed line). Also plotted
({\Large {\bf $\circ$}}) is $H_{N-1}^{\mathrm{min}}(T)/J$ for
$N$$=$$10$, the field corresponding to the minimum of the metastable
$|\mathrm{Im} f_\alpha|$ in the field interval $1$$<$$H/J$$<$$2$.}
\label{figb2}
\end{figure}


\end{document}